\documentclass[default,iicol]{sn-jnl}%

\usepackage{footmisc}

\begin{document}
\let\WriteBookmarks\relax
\def\floatpagepagefraction{1}
\def\textpagefraction{.001}

\title [mode = title]{Material parameter influence on the expression of Solitary-Wave-Induced Surface Dilation\footnotemark[2]}                      

\author*[1]{\fnm{Eric S.} \sur{Frizzell}}\email{efrizz@umd.edu}

\author[1]{\fnm{Christine M.} \sur{Hartzell}}

\affil[1]{\orgdiv{University of Maryland}, \orgname{Aerospace Engineering}, \orgaddress{\street{3178 Glenn L. Martin Hall}, \city{College Park}, \state{MD} \postcode{20740}, \country{United States}}}

\abstract{We formulate a method for predicting peak particle forces in a Solitary Wave (SW) wavefront within a randomly filled 3D granular channel. The SW in our simulation are driven by a sustained impact originating in the bumpy floor of the channel. We show that, when generated in this manner, forces in the driven SW wavefront within the 3D assembly follow the same power law scaling on material properties and impact velocity as in a 1D chain. A simple scaling of the 1D forces matches results from simulated impact tests we conduct using Soft Sphere Discrete Element method simulations. We then quantify the magnitude of Solitary Wave Induced Surface Dilation (SID) that occurs as a result of varied material properties and gravitational environments, giving an equation that can be used to predict the lofting depth (depth to which particles experience bulk density changes as a result of a laterally propagating SW wavefront). As predicted by our equation and confirmed with simulated results, SID is amplified as particle material properties become closer to lunar regolith grains, supporting the hypothesis that SID is the Lunar Cold Spot formation mechanism.}

\keywords{Inertial dilation, SSDEM, granular shock, Lunar Cold Spots}

\maketitle

\begingroup
\renewcommand\thefootnote{\fnsymbol{footnote}} 
\footnotetext[2]{This version of the article has been accepted for publication, after peer review but is not the Version of Record and does not reflect post-acceptance improvements, or any corrections. Version of Record: https://doi.org/10.1007/s10035-024-01460-0. Use of this version subject to the publisher’s terms: https://www.springernature.com/gp/open-research/policies/accepted-manuscript-terms.}
\endgroup

\section{Introduction} \label{sec:intro}

In this paper we first develop a method to predict the peak force felt by a particle in a randomly packed 3D granular assembly as it experiences a laterally propagating wavefront. Recent simulations by \cite{frizzell2023simulation} showed that impacts of a lateral piston within a shallow granular channel (exposed to vacuum and microgravity) led to a solitary wave (SW) which propagated over long distances. In this work, we consider briefly the driving mechanism leading to such long range propagation (the SW in \cite{frizzell2023simulation} was driven through constant impacts with the floor), but our focus is characterizing bulk dilation changes as a result of the SW, not its origin.  In the wake of the driven SW wavefront, a region of substantial bulk dilation near the surface is left behind as a result of particle lofting, a phenomenon that we call SW Induced Dilation, or SID. We are particularly interested in characterizing SW wavefront forces in 3D granular channels as our second task in this work is to understand the behavior of SID as a function of particle material properties. \cite{frizzell2023simulation} hypothesized that the impulse delivered to particles in the SW wavefront would be enhanced under lunar conditions (i.e. with smaller and harder grains), but this hypothesis relied on several assumptions. One important assumption was that particle velocities within a driven SW wavefront were constant regardless of grain material properties, but this is not the case (\cite{nesterenko2013dynamics}, eq. 1.41). Additionally, the forces that particles experience will also change as a function of their collisional velocity. Here, we will adapt a method used to calculate SW forces in a 1D granular chain (requiring only a simple rescaling) to predict the peak force experienced by a grain in our randomly packed 3D channels,  $F_{m,3D}$. We assess the appropriateness of our 3D SW wavefront force model by comparing predictions to simulated results. We generate driven SW in granular channels filled with particles of varied size, density, and modulus of elasticity under a range of impact velocities using Soft Sphere Discrete Element Method (SSDEM) simulations. Seeing that simulated $F_{m,3D}$ follows the expected power law scaling, we then construct a force balance between these driven SW wavefront forces and gravitational overburden, resulting in an equation that we solve to find the depth to which SID will occur ($z_{LD}$). We measure the SID response of our channels to the different SWs that we generate by quantifying the height change of the channel ($\Delta z$) as well as the bulk volume change as function of depth $\left(\% \Delta \rho(z)\right)$. Finally, we compare the simulated dilation to the  prediction and examine the sensitivity of SID to some important simulation parameters (friction, coefficient of restitution, etc). Our primary motivation is to understand how the magnitude of SID changes as particle hardness increases. A large dilation response for particles with lunar regolith-like properties has implications for SID as the formation mechanism of Lunar Cold Spots (LCS) \cite{bandfield2014lunarCS}, large distal regions of reduced surface bulk density surrounding craters on the Moon. Before discussing our simulation and results, we review a few important details of granular wave propagation, discuss how such a wave could lead to surface lofting of particles and consider possible implications for LCS formation.

\subsection{Granular SWs} \label{sec:granular_sw_p2}
Wave propagation in 1D granular chains is an extensively studied topic. In the most idealized case, a wave is initiated in a chain of spheres (particles/grains) by delivering an impulse to the first sphere. \cite{nesterenko1984propagation} first showed that grains at low or no pre-compression develop SW following a strong impact. SW travel in energy packets of roughly 5 particle diameters \cite{nesterenko2013dynamics} and when the coefficient of restitution of the particles is unity, no energy is lost while the packet propagates \cite{manciu2001impulse}. When the chain has the right amount of pre-compression (not strongly compacted, but also has some finite pre-compression) long lived SW arise (\cite{takato2012long}, \cite{jiao2023revisiting}). There are many studies that characterize the properties of SW in both 1D chains and 2D assemblies under varying conditions and there is an overview of the state of the art in \cite{frizzell2023simulation}, section 1.1. In 3D, SW excited via ground impact have even been used as a tool for detecting buried objects (\cite{rogers1994location}, \cite{sen1998solitonlike}, \cite{sen2005using}, \cite{visco2004impulse}). Here, we are most interested in studies that predict the force on particles within a laterally propagating SW wavefront.  \cite{sen1996sound} gave equations for the force on a grain in a vertically propagating wavefront, but this approach requires knowledge of the maximum displacement in the wavefront (and we are concerned with lateral propagation). Hasan gives equations for the force experienced by a grain in a 1D crystal as it experiences a SW (\cite{hasan2016universal}, \cite{hasan2017basic}, \cite{hasan2018shock}), but their model requires knowledge of the loading profile. Pal mostly considered elasto-plastic chain propagation (\cite{pal2013wave}), but they also considered a simple elastic Hertzian model similar to \cite{frizzell2023simulation}. Their equation for force in the wavefront depends only on the impact velocity and grain material properties (mass, elasticity, and size) \cite{pal2014characterization} and we will use this as the basis for predicting forces felt by a particle within a 3D driven SW wavefront ($F_{m,3D}$, see sec. \ref{sec:wavefront_forces}), which to the best of our knowledge has not yet been carried out. We propose that $F_{m,3D}$ can be approximated as a simple scaling of the peak forces in SWs propagating in 1D granular chains. We can then consider how changing forces in the wavefront influence the depth range over which surface particles experience lofting via SID and thus bulk density changes.

\subsection{SW-Induced Surface Dilation} \label{sec:SID}
We briefly review the SID mechanism leading to surface bulk dilation as it was initially described before providing some important clarifications on the process. \cite{frizzell2023simulation} used a lateral piston impact into a 3D granular channel to initiate SW. In those simulations, we described a SW that develops following the brief existence of a planar shock wave. The SW then maintained its strength and shape as it propagated over meter scale distances. As a result of proximity to vacuum, the speed of this wave in the near-surface increases quickly as overburden pressure increases with depth, leading to a curvature in the wavefront as the SW propagates. The SW travels just above the local sound speed, which similarly increases with depth. As particles experience the SW wavefront they undergo a frictional collision with their neighbors and develop an upwards velocity. The curvature of the wavefront ensures this upward motion and the result is an entire sheet of lofted particles on ballistic trajectories. SID also requires a granular assembly to be in a compacted state in order for particle lofting to occur; below a threshold packing fraction, the assembly will compact instead of dilate after experiencing a SW. Particles are only able to loft when the force imparted in the vertical direction by the SW wave exceeds the force particles feel from the overburden and this is wavefront force is where $F_{m,3D}$ will be used. In \cite{frizzell2023simulation}, the maximum speed of the SW occurred along a rough channel floor and was determined by the piston impact velocity.  Since the largest velocity occurs at the floor, the effect of SID is reduced in taller channels (i.e., when surface grains are further from the floor). For the same piston impact speed the wavefront curvature evolves over a greater vertical distance which results in a slower traveling SW at the surface. Collisional velocity is therefore reduced as is the vertical velocity near-surface grains can attain. 

We clarify here that the wave generation process used in \cite{frizzell2023simulation} actually produces two distinct SW, not a shock that decays to a steady SW as initially introduced. The piston impact generates a SW (PSW) which disperses as it propagates while at the same time a second SW (which ultimately overtakes the first) is driven by continual impacts propagating along and within the rough channel floor. The same velocity that acts as an impulse for virtual piston particles is also transmitted through our thin sheet of floor particles and, since the floor is made up of randomly packed particles, the un-fixed grains of the channel in contact with the fixed floor experience oblique impacts. These impacts drive the constant-strength driven SW (DSW) across the channel and the work in \cite{frizzell2023simulation} (and here) represent an idealized scenario of `perfect-floor impact' (discussed further in sec. \ref{sec:wavegen}). Without the DSW, the energy of the piston impact would decay through our granular channel over a finite length. In fact, grains are often used as a type of force absorption \cite{sen2017impact} given their ability to efficiently decimate propagating energy. Any wave traveling over a significant distance through small grains (e.g., 1 mm or less like in \cite{frizzell2023simulation}) would require an external driver. This work does not seek to explain that external driver and we continue to use the idealized `perfect-floor impact' to study SW via the DSW. Our focus in this work is to understand how the reaction of near-surface grains to the driven SW changes as the material properties of grains in the channel become regolith-like.

\subsection{Lunar Cold Spots}
LCS are apparent as relatively cold halos surrounding source craters on the Moon. Their low thermal inertia is hypothesized to be the result of bulk density differences down to 40 cm in depth. While the LCS appear around all young craters (\cite{williams2018lunarCS}) and can extend more than 100 radii from the source, there are no visible signs of surface modification. The LCS halo becomes apparent around 10 crater radii from the source crater, beyond the region where impact ejecta is evident. The interested reader can refer to the LCS literature for further details: \cite{bandfield2014lunarCS}, \cite{williams2018lunarCS}, \cite{hill2017well}, \cite{venkatraman2023statistical}, \cite{powell2023high}. In \cite{frizzell2023simulation} we hypothesized that SID was a possible LCS formation mechanism given the SW produced bands of near surface dilation that appear to be consistent with expected LCS bulk density changes. However, the material properties of our granular assembly did not represent lunar surface regolith and it is unclear how SID will scale to assemblies with different material properties. Understanding how forces in the SW wavefront ($F_{m,3D}$) scale with material parameters will provide further evidence for SID as the LCS formation mechanism.

\begin{figure}[h]
\begin{center}
\includegraphics[scale = .50]{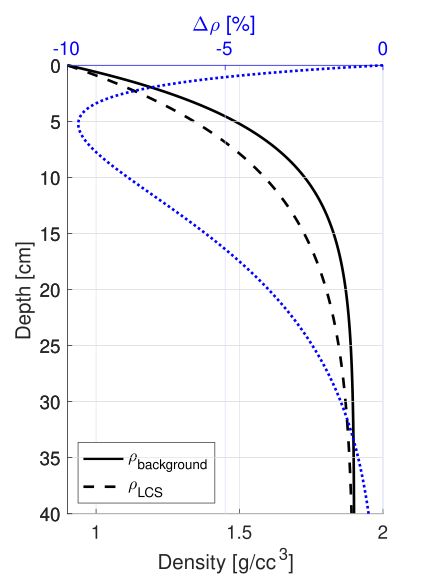}
\end{center}
\caption{\textbf{Cold Spot density vs depth.} The lower X axis represents bulk density (black) and the upper X axis represents the percent change in bulk density density (blue), where negative percent change represents dilation. The y axis shows depth in cm and is the shared by both trends. The lunar surface regolith (black solid line) and LCS (black dashed line) are predicted using the method and fit parameters from \cite{bandfield2014lunarCS}. Bulk density change (dotted blue line) takes $\rho_{LCS}$ to be a percent difference from $\rho_{background}$. This percent difference can be thought of as `dilation expected as the result of LCS formation' and we will use it as a reference later when viewing Figs. \ref{fig:dilation_vs_depth_phi} and \ref{fig:dilation_vs_depth_combo}.}
\label{fig:bandfield_density_pdiff}
\end{figure}

In this paper, we will compare bulk density at the surface and at depth to expected density changes for typical LCS. Figure \ref{fig:bandfield_density_pdiff} shows the hypothesized bulk density profile for background and LCS surface regolith based on Diviner thermal measurements \cite{bandfield2014lunarCS}. Density differences are likely constrained to the first 40 cm of surface regolith and are substantial near the surface. Evaluating the percent change between background and LCS regolith, the peak density change of about 10\% would occur around 5 cm in depth and the average density change over the entire depth region between LCS and background regolith is about 5\%. If SID is involved in LCS creation, we'd expect our simulated SID response to exceed the average expected density change since the channels in our simulations are only a fraction of the total LCS depth (in other words, our channel is shorter than the lunar `channel' and a shorter channel leads to greater wave speeds - sec. \ref{sec:SID}).

\section{Methodology}  \label{sec:Methodology_p2}
We discuss our computational model, a method for estimating SW wavefront forces using a 1D chain assumption, and a method for predicting the depth to which particles experience lofting following the passage of a SW. We provide details on our simulation and tests performed, computational resources and data analysis methods.

\subsection{Simulation setup} \label{sec:SimSetup_p2}
Our simulation implementation is largely the same as \cite{frizzell2023simulation} which used a particle piston to initiate waves within a long, thin granular channel filled with a monodisperse assembly of grains under lunar gravity. The grains are confined by end walls and a rough floor but are subject to periodic boundary conditions in the azimuthal direction which allows our simulated channel to be representative of a volume element within the path of a radially expanding pressure wave. After the channel experiences a lateral impulse delivered by a virtual piston with impact velocity $v_0$, the PSW decays in strength until the DSW becomes dominant. The maximum $v_0$ considered corresponds to a fixed portion of $c_m$ (roughly 20\% $c_m$). With the material parameters used in \cite{frizzell2023simulation} this gives a maximum impact piston velocity $v_0$ =  10 m/s. We use the same limit here. We select a single channel geometry of 2 meters in length, filled 20 cm deep, and 2 cm wide. For the base case particle size of radius $R$ = 1.25 mm, this fixes the size of our channel (in terms of particle radii) at 1600, 160, and 16 (length, height, width respectively). The decay range and wall effect regions of the channel are usually a function of particle number ($\sim$30 particles in a 1D chain, \cite{nesterenko2013dynamics}) so we can therefore keep the same channel geometry while scaling particles size or material parameters and still have a region of the channel which provides an unadulterated view of DSW effects.

We again use \textit{LIGGGHTS} \cite{kloss2012models}, an open source Soft Sphere Discrete Element Method (SSDEM) solver, to perform our simulations. SSDEM is a common methodology used to study low gravity granular dynamics (\cite{sanchez2011simulating}, \cite{schwartz2012implementation}, \cite{tancredi2012granular}, \cite{sanchez2016disruption}, \cite{demartini2019using}, \cite{zhang2021creep},  \cite{sanchez2022transmission}), but comes with a potentially large computational cost (depending on particle properties) due to integrating the equations of motion at a small time step ($<$1\% of Hertz collision time $\tau_c$, eq. \ref{eq:contact_time_p2}). We can reduce compute time in many cases by re-using the particle channels that were prepared in \cite{frizzell2023simulation}, but varying particle size (varied $R$) requires additional channel preparation.

\subsubsection{Pre-settled channels} \label{sec:prefilled}
We use (from \cite{frizzell2023simulation}) 2 meter long channels filled 20 cm deep with a monodisperse particle distribution in three packing configurations  (loose\footnote{Filename restart2m20cm\_loose.static}  $\phi_0\sim$55\%, medium\footnote{Filename restart2m20cm\_medium.static}  $\phi_0\sim$59\% and compact\footnote{Filename restart2m20cm\_compact.static}  $\phi_0\sim$62\% channels found at \cite{frizzell2023data}), where $\phi_0$ = $V_{particles}$/$V_{channel}$ is the initial packing fraction. The base material parameters of the particles are elastic modulus $E$ = 5 MPa, particle density $\rho$ = 2.5 g/cc (note that any equations involving density should use SI units, kg/m$^3$)  particle radius $R$ = 1.25 mm, cohesion energy density $k_c$ = 1 kPa, Poisson ratio $\nu$ = 0.2, sliding friction $\mu =$ 1, rolling friction $\mu_r$ = 0.8, rolling viscous damping $\gamma_{d,r}$ = 2 and restitution $e$ = 0.5. We will often refer to these material and simulation parameters as the base properties. To investigate how our driven SW propagation is altered by material properties, we take the base case channel, change the parameter of interest, and allow a settling period to reduce background forces before saving the new channel state. The settled channel now has the new material parameters and can be used multiple times (creating multiple SW from different $v_0$). Given the form of eq. \ref{eq:contact_time_p2}, altering particle density, size, and modulus all influence the contact time which has implications for our simulation step size. $\nu$ also influences $\tau_c$, but the range of $\nu$ we investigate is small compared to the order of magnitude sweeps we will conduct in E, $\rho$ and R space.  The majority of our simulations have a time step set to $<$1\%$\tau_c$ which we show in Tables \ref{tab:E_dt_check} and \ref{tab:rho_dt_check}. A few cases (like $E$ = 50 MPa, $v_0$ = 31.6 m/s) slightly exceed the 1\% limit, though we determined that to be acceptable since results were consistent with simulations with a time step $\ll$1\%$\tau_c$. Additionally, the time step is set using the maximum $v_0$ which occurs at t = 0s when the piston is initiated as well as along the floor as the `perfect floor impact' velocity travels through particles in the rough base layer. Our time step is smaller than needed in most cases since the surface lofting we are interested in occurs where particle velocities are slower. Tables \ref{tab:E_dt_check} and \ref{tab:rho_dt_check} also show the elastic limit, $L_e$, which is the collision speed above which Hertz theory begins to break down. $L_e$ will be used later (Table \ref{tab:validity}) to discuss the validity of particle collision speeds within the wavefront.

One thing to keep in mind regarding the channel preparation method discussed here is that it is distinctly different from the pouring method used in the next section. Preparation history of the channel does influence channel response and this can be seen later in sec. \ref{sec:wavefront_force_fitting} when changing $\rho$ using the `pre-settled' method leads to looser channels (smaller $\delta_0$ in the varied $\rho$ plot of Fig. \ref{fig:dm_and_d0_vs_depth}) than expected. However, we deem these differences to be negligible given the agreement we will later see with expected behavior (scaling of wavefront forces and surface dilation) across multiple decades of parameter space as well as between channels prepared using both preparation methods (see secs. \ref{sec:wavefront_force_fitting} and \ref{sec:dilation_comparison}). 

\begin{table*}[h]
\begin{minipage}{430pt}
\caption{Channel properties for testing sensitivity to elastic modulus. We give the average initial conditions, velocity limits, and assess the appropriateness of our time step choice in channels of varying $E$. $\delta_0$ decreases as E increases since the spring repulsion between particles increases and they push apart, resulting in a slightly looser channel (smaller $\phi_0$). The largest $\delta_0$ (experienced by the particles with $E$ = 5 MPa) is about 0.3\% of $R$ which is sufficiently small. $\Delta t$ is well below 1\% of $\tau_c$ for most cases, though this limit is exceeded slightly in few of the cases (up to 1.46\% of $\tau_c$). We determined that these tests were still run at a sufficiently small time step given the agreement of the behavior of the SW in theses channels with expectation (sec. \ref{sec:wavefront_force_fitting}). Other particle parameters are constant ($R$ = $1.25$ mm, $\rho$ = $2.5$ g/cc). We performed two series of tests in these channels, one with $v_0$ constant at 10 m/s and one where $v_0$ was a constant portion of $c_m$ ($\sim$20\% $c_m$). A dashed line indicates `same as above'.}\label{tab:youngs}

\begin{tabular}{@{}ccccccccc@{}}
\toprule
\begin{tabular}{@{}c@{}} E \\ (MPa) \end{tabular} & 
\begin{tabular}{@{}c@{}} $\phi_0$ \\ (\%) \end{tabular} & 
\begin{tabular}{@{}c@{}} $\delta_0$ \\ ($\mu$m) \end{tabular} & 
\begin{tabular}{@{}c@{}} $c_m$ \\  (m/s) \end{tabular} & 
\begin{tabular}{@{}c@{}} $v_{0}$ \\ (m/s) \end{tabular} & 
\begin{tabular}{@{}c@{}} $\tau_c$ \\ (s) \end{tabular} & 
\begin{tabular}{@{}c@{}} $\Delta$ t \\ ($s$) \end{tabular} & 
\begin{tabular}{@{}c@{}} $\Delta$ t/$\tau_c$ \\ (\%) \end{tabular} &  
\begin{tabular}{@{}c@{}} $L_e$ \\ (mm/s) \end{tabular} \\
\midrule
5 & 62.08 $\pm$ 0.048 & 4.23 $\pm$ 1.50  & 47.1 & 10.0 &  $2.166\times10^{-4}$ & $10^{-6}$ & 0.46  & 9.4 \\
\hline
20 & 61.89 $\pm$ 0.051 & 2.18 $\pm$ 8.81  & 94.2 & 20  &  $1.733\times10^{-4}$ & $10^{-6}$ & 0.92 & 18.9  \\
- & - & - & - & 10.0  &  $1.990\times10^{-4}$ & - & 0.81   & -  \\
\hline
50 & 61.71 $\pm$ 0.055 & 1.16 $\pm$ 0.58 & 149.1 & 31.623 &  $6.848\times10^{-4}$ & $10^{-6}$ & 1.46 & 29.81  \\
- & - & - & - & 10 &  $8.622\times10^{-5}$ & - & 1.15 & -  \\
\hline
200 & 61.63 $\pm$ 0.051 & 0.752 $\pm$ 0.512 & 298.1 & 63.24  &  $3.424\times10^{-5}$ & $10^{-7}$ & 0.29 & 59.63   \\
- & - & - & - & 10.0  &  $4.952\times10^{-5}$ & - & 0.20 & -   \\
\hline
500 & 61.55 $\pm$ 0.051 & 0.384 $\pm$ 0.312 & 471.4 & 100.0 &  $2.166\times10^{-5}$ & $10^{-7}$ & 0.46 & 94.28  \\
- & - & - & - & 10.0 &  $3.432\times10^{-5}$ & - &0.29 &  -  \\
\hline
5,000 & 61.54 $\pm$ 0.049 & 0.046 $\pm$ 0.018 & 1490.7 & 316.23 &  $6.848\times10^{-6}$ & $10^{-7}$ & 1.46 & 298.14 \\
- & - & - & - & 10 &  $1.366\times10^{-5}$ & - & 0.73 & -   \\

\bottomrule
\end{tabular} \label{tab:E_dt_check}
\end{minipage}
\end{table*}

\begin{table*}[h]
\begin{minipage}{430pt}
\caption{Channel properties for testing sensitivity to density. We give the average initial conditions, velocity limits, and assess the appropriateness of our time step choice in channels of varying $\rho$. The largest $\delta_0$ (experienced by the particles with $\rho$ = 25 g/cc) is about 1\% of $R$ which is approximately the limit used in \cite{schwartz2012implementation}. We judged this large overlap to be acceptable since most cases we evaluated use smaller $\rho$. $\delta_0$ increases as $\rho$ increases since, as a result of increased mass, the particles have a reduced spring repulsion force. The particles are able to interpenetrate further as $\rho$ increases, leading to more `compact' (larger $\phi_0$) channels. $\Delta t$ is well below 1\% of $\tau_c$ for all cases. Other particle parameters and impact velocity are constant ($R$ = 1.25 mm, $E$ = 5 MPa, $v_0$ = 10 m/s).}\label{tab:particle_density_timestep}
\begin{tabular}{@{}cccccccc@{}}
\toprule
\begin{tabular}{@{}c@{}} $\rho_{p}$ \\ (g/cm$^3$) \end{tabular} &
\begin{tabular}{@{}c@{}} $\phi_0$ \\ (\%) \end{tabular} & 
\begin{tabular}{@{}c@{}} $\delta_0$ \\ ($\mu$m) \end{tabular} & 
\begin{tabular}{@{}c@{}} $c_m$ \\  (m/s) \end{tabular} & 
\begin{tabular}{@{}c@{}} $\tau_c$ \\ (s) \end{tabular} & 
\begin{tabular}{@{}c@{}} $\Delta$ t \\ (s) \end{tabular} &
\begin{tabular}{@{}c@{}} $\Delta$ t/$\tau_c$ \\ (\%) \end{tabular} &
\begin{tabular}{@{}c@{}} $L_e$ \\ (mm/s) \end{tabular} \\
\midrule
$0.8$ & 61.94 $\pm$ 0.05 & 2.54 $\pm$ 0.96 & 83.3 & $1.373\times10^{-4}$ & $10^{-6}$  & 0.73 &  16.7 \\
$1.5$ & 62.03 $\pm$ 0.05  & 3.48 $\pm$ 1.20 & 60.9 & $1.765\times10^{-4}$ & $10^{-6}$ & 0.57 &12.2  \\
$2.5$ & 62.08 $\pm$ 0.05  & 4.23 $\pm$ 1.50 & 47.1 & $2.166\times10^{-4}$ & $10^{-6}$  &0.46 &9.4  \\
$5.0$ & 62.20 $\pm$ 0.05  & 5.78 $\pm$ 2.20 & 33.3 & $2.858\times10^{-4}$ & $10^{-6}$ & 0.35  &6.7 \\
$10.0$ & 62.40 $\pm$ 0.05  & 8.40 $\pm$ 3.38 & 23.6 & $3.771\times10^{-4}$ & $10^{-6}$ & 0.27 &4.7  \\
$15.0$ & 62.53 $\pm$ 0.05  & 10.62 $\pm$ 4.38 & 19.2 & $4.434\times10^{-4}$ & $10^{-6}$ & 0.23  & 3.8 \\
$25.0$ & 62.93 $\pm$ 0.07  & 14.27 $\pm$ 6.00 & 14.9 & $5.440\times10^{-4}$ & $10^{-6}$ & 0.18 &3.0  \\

\bottomrule
\end{tabular} \label{tab:rho_dt_check}
\end{minipage}
\end{table*}

\subsubsection{New channel filling} \label{sec:variedR_fill}

The particle sizes used in sec. \ref{sec:prefilled} are roughly an order of magnitude larger than lunar grains. The large role of cohesion on lunar grain dynamics (\cite{hartzell2011role}) is one example of unique behavior that emerges when particles are relatively small. Since it is not clear how the strength of the SW will change with particle size, we explore both larger and smaller grains than in \cite{frizzell2023simulation}.  When changing particle size it is necessary to re-pour the channels in order to maintain the same $\phi_0$. To do this, we scale the filling procedure from \cite{frizzell2023simulation} by the ratio of the newly investigated particle radius to the base case $R$ = 1.25 mm (ratios shown in Table \ref{tab:particle_size_timestep}). We used a medium filling (sec. \ref{sec:prefilled}) to reduce computation time. To be able to compare SID results across the various particle sizes, we scale the dimensions of our channels by the same factor to maintain a fixed height in length in terms of number of particles (length scales are non dimensionalized by particle radius in this work). We repeat the time step checks for varied $R$ in Table \ref{tab:particle_size_timestep}. All of the channels with varying grain size have a consistent $\phi_0$ (just below 59\%) except for the 12.5 mm radius particles which filled slightly looser. This is a result of the scaling procedure, where the particle input velocity is increased 10x over the 1.25 mm radius case but $c_m$ does not change since it is not dependent on $R$. This means that, for this case only, particles interpentrated more than is physically possible as they impacted the channel during pouring. When the pouring is completed the particles find themselves in a `too compacted' state and undergo rearrangement to a looser state. Given the consistency in the other particle size channels, we determined that the difference in $\phi_0$ (about 2\% less) for the 12.5 mm radius is acceptable though it should be considered when analyzing resultant SW forces and the SID response in the channel. 

We emphasize here that our 80 particle channel height ($\sim$160$R$) is only a small fraction of the LCS 40 cm depth. Depending on the size used for lunar surface grains (tens or hundreds of microns), the LCS depth range corresponds to 1,000 to 2,000 particle diameters. Therefore in this paper we are only examining a small segment ($\sim$1\%) of the LCS channel. One thing to keep in mind is that the reduction in height of our channel compared to reality means that, for the same $v_0$ that we consider in our 160$R$ channels, particles near the surface of a taller channel would experience smaller near-surface forces (\cite{frizzell2023simulation}) and thus see a reduction in the magnitude of SID experienced. Furthermore, the fixed channel height means that smaller particles experience smaller overburden forces than larger particles. If SID is involved in LCS creation, our simulations should over predict dilation compared to the expectation in Fig. \ref{fig:bandfield_density_pdiff} for the assumed grain properties.

\begin{table*}[h]
\begin{minipage}{425pt}
\caption{Channel properties for testing sensitivity to particle size. Particle properties are set with to base case values ($E$ = 5 MPa, $\rho$ = 2.5 g/cm$^3$, $v_0$ = 10 m/s). We pour the particles to a medium packing fraction (around 58\%) which was chosen as a balance between reducing channel preparation time while achieving a level of $\phi_0$ that would still result in SID. We take the elastic limit to be $L_e$ = 9.43 mm/s for all cases since $c_m$ does not depend on $R$.} \label{tab:particle_size_timestep}
\begin{tabular}{@{}ccccccccc@{}}
\toprule
\begin{tabular}{@{}c@{}} $R$ \\ (mm) \end{tabular} &  
\begin{tabular}{@{}c@{}} $\phi_0$ \\ (\%) \end{tabular} & 
\begin{tabular}{@{}c@{}} $\delta_0$ \\ ($\mu$m) \end{tabular} & 
\begin{tabular}{@{}c@{}} $\tau_c$ \\ (s) \end{tabular} &  
\begin{tabular}{@{}c@{}} $\Delta$ t \\ (s) \end{tabular} & 
\begin{tabular}{@{}c@{}} $\Delta$ t/$\tau_c$ \\ (\%) \end{tabular} &  
\begin{tabular}{@{}c@{}} $L$ \\ (m) \end{tabular} &  
\begin{tabular}{@{}c@{}} $H$ \\ (cm) \end{tabular} &  
\begin{tabular}{@{}c@{}} Scale \\ (\#) \end{tabular}  \\
\midrule
$0.125$ & 58.61 $\pm$ 0.067 & 0.0996 $\pm$ 0.0359 & $2.166\times10^{-5}$ & $10^{-7}$ & 0.46 & 0.2 & 2 & 0.1  \\
$0.50$ & 58.87 $\pm$ 0.066 & 1.027 $\pm$ 0.368 & $8.663\times10^{-5}$ & $10^{-7}$ & 1.15 & 0.8 & 8 & 0.4   \\
$1.25$ & 58.84 $\pm$ 0.046 & 4.810 $\pm$ 1.822 & $2.166\times10^{-4}$ & $10^{-6}$ & 0.46 & 2 & 20 & 1 \\
$5$ & 58.26 $\pm$ 0.083 & 53.40 $\pm$ 21.57  & $8.663\times10^{-4}$ & $10^{-6}$ & 0.12 & 8 & 80 & 4 \\
$12.5$ & 56.29 $\pm$ 0.166 & 208.0 $\pm$ 90.86  & $2.17\times10^{-3}$ & $10^{-6}$ & $0.046$  & 20 & 200 & 10 \\
\bottomrule
\end{tabular}
\end{minipage}
\end{table*}

\begin{figure}[h]
\begin{center}
\includegraphics[scale = .42]{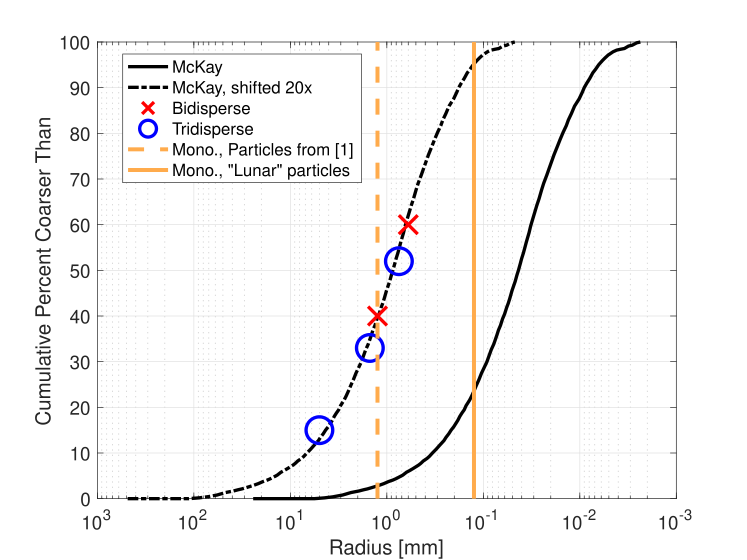}
\end{center}
\caption{\textbf{Particle size distributions.} We plot cumulative percent of particles greater than a given size vs particle radius (mm). We use the lunar surface particle distribution from Fig 7.9 of \cite{mckay1991lunar} (data harvested with \cite{Rohatgi2022}) corresponding to the 63\% Agglutinate soils (solid black line). The black dashed line is this same trend from \cite{mckay1991lunar}, but increased by just over one order of magnitude (20x). The dashed line is the distribution from which we determine our bi-disperse (red Xs) and tri-disperse (blue circles) mixtures. The dashed orange line shows the particle size for the base case monodisperse distribution used in this paper (and in \cite{frizzell2023simulation}) while the solid orange line shows the size distribution for the monodisperse `lunar-like' particles used in this paper.}
\label{fig:particle_distributions}
\end{figure}

We also inspect how polydispersity influences the ability of the SW to dilate surface regolith. Dispersity tends to increase dilation responses (\cite{jia1999ultrasound}) and so we expect the same to be true for SID. The distributions for the bi- and tri-disperse mixtures we considered are given in Fig. \ref{fig:particle_distributions} and are shown alongside the distribution of lunar surface particles from \cite{mckay1991lunar}. Though we used the same filling procedure for these channels as the base case, introducing dispersity decreased $\phi_0$ resulting in $\phi_0$ = 54.39 $\pm$ 0.03 and  52.46 $\pm$ 0.21 for the bi and tri disperse channels (respectively).

\subsubsection{Wave generation} \label{sec:wavegen}

Following either filling procedure, the particle states are saved so that channel preparation is not required for every SW test. After the settled channel state is loaded, we freeze a horizontal sheet of particles to the floor to ensure a rough bottom. We then give particles in a vertical sheet within one particle diameter of the left wall a lateral velocity $v_0$ and they act as a piston to initiate the PSW. As a result of our rough floor and piston implementation, $v_0$ is also transmitted through the floor particles. Although floor particles are `frozen' in the simulation and do not move, their velocity information is updated (the force on the floor particles is set to zero before integrating the equations of motion, but the velocity of the contact is stored separately and is passed to the next particle). When a region of the channel particles in contact with frozen floor particles experiences this `frozen' velocity, they experience what is essentially an impact. The DSW is driven across the channel by continuous floor impacts. We highlight that the impacts between channel particles and floor particles are oblique; the initial $v_0$ frozen into the floor propagates laterally. We refer to the length of the channel over which the PSW is active as the blast loading (BL) region. By evaluating the average particle overlap $\delta$, we can characterize dissipation of the PSW (in the BL region) and the strength of the driven DSW.

In the DSW region, the wavefront propagates with constant strength as a result of the sustained floor impacts. Fig. B.1 from \cite{frizzell2023simulation} shows a comparison between the wavefronts of the initial piston impact (frame b) and the DSW (frames c and d). The initial wavefront is uniform and flat, but the DSW wavefront is curved as a result of increasing sound speed with depth. The shape of the front is affected by material parameters; harder particles and larger $v_0$ lead to more curvature near the surface while increasing particle size alone results in a shallower depth range over which curvature exists. For $v_0$ that are below $c_0$, the decay of the PSW wavefront can be seen as it traverses the channel since the DSW propagates more slowly in this case. The frozen in $v_0$ still propagates along the floor of the channel and, instead of generating a DSW, periodically delivers an impulse of sufficient magnitude to perturb particles. When particles are `soft' (lower $E$), these periodic outbursts do not cause surface dilation, but harder particles ($E$ $>$ $\sim$ 500 MPa) experience substantial dilation, at times nearly as much as in the case of a DSW wavefront. We will revisit these lower speed floor impacts in sec. \ref{sec:floor_vortex}, but here we emphasize that the focus of this work (and therefore $F_{m,3D}$) is to characterize forces in the driven SW wavefront, not the outbursts.

\subsection{Contact model} \label{sec:model}
We use the same approach as \cite{frizzell2023simulation} to model the grains and collisions within our granular assembly (see the appendix in \cite{frizzell2023simulation}). We briefly summarize the model here. Our particles are Hertzian spheres subject to friction (sliding, rolling, twisting) and cohesion. Forces in the normal and tangential (subscripts $n$ and $t$) direction of the contact frame are given in equation \ref{eq:force_model}. The Hertzian spring constant is $k$ $\left((4/3)ER^{1/2}\right)$, particle overlap is $\delta$, $v$ is the relative velocity between two colliding particles, $\gamma$ is a viscous damping term which depends on material parameters and coefficient of restitution $e$, $k_c$ is a cohesion energy density, $A_c$ is the contact area, and $\mu$ is the friction coefficient. Particles are also subject to a downward uniform gravitational field. 

\begin{align}
    \textbf{F}_n = - k_n \delta_n^{3/2} - \gamma_nv_n + k_cA_c \nonumber  \\
     \textbf{F}_t = - k_t \delta_t^{3/2} - \gamma_tv_t  \nonumber  \\ \mid F_t \mid \leq \mu_s \mid F_n  \mid \label{eq:force_model}
\end{align}

Using this model we can define several relevant quantities related to wave propagation in the channel. When the channel is in a quasi-static state (pre-impact), a particle at depth will experience a fixed gravitational load based on the weight of particles above it (overburden). Approximating a 3D granular assembly as a 1D stack of particles, we can approximate how $\delta_0$ (initial overlap) changes with depth. We approximate the weight of the overburden as in equation \ref{eq:overburden}, where $\rho$ is the particle density. The first term represents the bulk density, the second term is the volume of a column of particles resting on top of a single particle at depth $z$, and $g$ is gravitational acceleration.

\begin{equation}\label{eq:overburden}
    F_0 = \rho \phi \times z\pi R^2 \times g
\end{equation}

Assuming the stack is quasi-static ($v$ is negligibly small), neglecting cohesion, setting the overburden force equal to the normal component of eq. \ref{eq:force_model}, and solving for $\delta_0$ leads to the approximate overlap as a function of depth in eq. \ref{eq:delta0}. We can evaluate this equation to get an idea of how overlap scales with material properties. Inserting eq. \ref{eq:overburden} into eq. \ref{eq:delta0} shows that $\delta_0$ is proportional to $E^{-2/3}$, $\rho_p^{2/3}$, $R$, and $z^{2/3}$. Neglecting cohesion is acceptable for the relatively low level of particle-particle cohesion used in most of our simulations and has very little influence on $\delta_0$, though we revisit this assumption later when considering Fig. \ref{fig:dm_and_d0_vs_depth}.

\begin{equation}\label{eq:delta0}
    \delta_0 = \frac{2\left(\frac{3(1-\nu^2)}{4E}F_0\right)^{2/3}}{R^{1/3}}
\end{equation}

Knowing the initial overlap, we can also compute the expected sound speed in the assembly. Continuing to use a 1D chain approximation, we give the sound speed $c_0$ as eq. \ref{eq:sound_speed_nesterenko} which is eq. 1.7 from \cite{nesterenko2013dynamics} modified with a $\sqrt{3}/2$ scaling as was done in \cite{frizzell2023simulation}.

\begin{equation}\label{eq:sound_speed_nesterenko}
    c_0 = \frac{\sqrt{3}}{2} \left(\frac{E(2R)^{1/2}}{3(1-\nu^2)m} \delta_0^{1/2}6R^2\right)^{1/2}
\end{equation}

Inserting $\delta_0$ and replacing the particle mass $m$ with $\rho V$ ($V = \frac{4}{3}\pi R^3$), sound speed as a function of depth is proportional to $E^{1/3}$ and $\rho^{-1/3}$. Notably, $c_0$ does not change with particle size when we consider $F_0$ purely as function of $z$. However, in this work we will report an average sound speed which is taken as the average of $c_0$ at each depth within a channel of particles.  Since we fix channel height based on number of particle radii (160$R$) we introduce an additional factor of $R$ into eq. \ref{eq:overburden} and we should therefore expect $c_0$ to scale with $R^{1/6}$ (slight increase with sound speed as particle size increases for our fixed height channels). We also point out that sound speed in a granular assembly ($c_0$) is much less than the sound of speed within an individual grain which is given as the material sound speed ($c_m$) in eq. \ref{eq:materialsoundspeed} (\cite{mase2009continuum}).

\begin{equation}\label{eq:materialsoundspeed}
    c_m = \sqrt{\frac{(1-\nu)E}{(1+\nu)(1-2\nu)\rho}} \\
\end{equation}

As in \cite{frizzell2023simulation}, we will limit particle impact velocities to less than 20\% of $c_m$. The material sound speed is proportional to $E^{1/2}$ and $\rho^{-1/2}$. Lastly, an important parameter in selecting an appropriate simulation time step is the Hertz contact time, given in equation \ref{eq:contact_time_p2}.

\begin{equation}\label{eq:contact_time_p2}
    \tau_c = 2.94 \left (\frac{M^2}{\left (\frac{16}{15} \right )^2 E_{eq}^2 R V_{in}} \right )^{1/5}
\end{equation}

\subsection{SW wavefront particle force} \label{sec:wavefront_forces}

The DSWs investigated in this work propagate laterally within a long granular channel filled randomly with particles. The randomness of the packing leads to highly nonlinear force propagation compared to the linear propagation along a 1D chain. While the increase in dimensionality leads to different dynamical response than in the 1D case, 3D systems do often exhibit the same phenomenon as 1D systems. \cite{el2008acoustic} saw order a two order of magnitude difference in wave speeds between 1D and 3D systems and the power law relationship of shock wave speed to impactor speed is recreated in both geometries (\cite{frizzell2023simulation}, \cite{gomez2012shocks}, \cite{van2013shock}, \cite{tell2020acoustic}). In this work, we will characterize the behavior of our 3D wavefront by approximating it as a 1D chain. The particles at depth $z$ are thought of as belonging to 1D granular chain which is subject to a finite compression (eq. \ref{eq:delta0}) induced by the overburden pressure at that depth. Since we are ultimately interested in understanding how the particle lofting behavior of SID is altered under different conditions, we need a method for predicting the force felt by a particle.

We can use conservation of energy to arrive at an expression for the force experienced by a particle in a SW wavefront. The peak force on the particle in the leading wavefront of a 1D chain can be solved for by equating the energy of a single grain to the kinetic energy of an impactor particle (\cite{awasthi2012propagation}). Setting the peak force in the chain ($F_p$) equal to the Hertz spring energy ($\frac{2}{5}k\delta_{p}^{5/2}$) can be solved to find the peak overlap $\delta_p = (F_{p}/k)^{2/3}$. Note that for our chosen simulation methodology (sec. \ref{sec:SimSetup_p2}) the maximum overlap should not exceed $\sim$1\% of the particle radius (\cite{schwartz2012implementation}). Taking the kinetic energy of an impactor grain to be $\frac{1}{2}mv_0^2$ and isolating $F_p$,  eq. \ref{eq:Pal_FSW} appears. \cite{pal2014characterization} found that the coefficient of 0.719 in eq. \ref{eq:Pal_FSW} is required to predict the force in the leading SW wavefront of a 1D chain.

\begin{equation} \label{eq:Pal_FSW}
    F_p = 0.719(m^3E_{eq}^2R^*v_0^6)^{1/5} 
\end{equation}

Arriving at this expression requires several assumptions including that all the energy of the collision goes into the SW and that the energy in the chain is exclusively in the wavefront. Eq. \ref{eq:Pal_FSW} also assumes that there is no initial pre-compression on the grains ($\delta_0 = 0$). While that is not the case for the channels we use in this paper (there is some small pre-compression due to gravitational loading), we are going to continue to use $\delta_0$ = 0 in formulating $F_{m,3D}$ assuming that the forces generated by the SW wavefront greatly exceed forces on a particle due to $\delta_0$. Similarly, we neglect the cohesive and damping terms of our model (eq. \ref{eq:force_model}) since the spring force should again be much greater than any damping or cohesive force contribution. We are also assuming that the 1D equation which is formulated for linear force transmission can be applied to the case of a 3D SW driven by oblique impacts along the floor. These simplifying assumptions will prove to be justified later in sec. \ref{sec:dilation_results}, but should be relaxed in future works. We proceed using the same form of eq. \ref{eq:Pal_FSW}, assuming that the forces in the 3D assembly will follow the same power scaling based on material parameters and $v_0$ as in 1D. We modify eq. \ref{eq:Pal_FSW} and write eq. \ref{eq:Fm_scaled} for $F_{m,3D}$ and will attempt to determine a numerical leading coefficient $C_1$ for a grain within the wavefront of a 3D SW. Note that eq. \ref{eq:Fm_scaled} reduces to the same form as eq. \ref{eq:Pal_FSW} when the definition of $k$ is inserted. In our simulations, we set the particle density as opposed to mass so the expected power law scaling (after inserting $m$ = $\rho\times$Volume) is $F_{m,3D}\sim$ $\rho^{3/5}$, $E^{2/5}$, $v_0^{6/5}$ and $R^2$.

\begin{align} \label{eq:Fm_scaled}
    F_{m,3D} = C_1 k  \left( \frac{5}{4}mv_0^2k^{-1}\right)^{3/5}
\end{align}

In this work we embark on a campaign of numerical simulations using varied material parameters to determine an appropriate value for $C_1$. We adjust one of the parameters that determine the force ($R$, $E$, $\rho$, $v_0$) and conduct a series of simulations with the adjusted parameter. Each test results a single average force produced by the SW for that parameter configuration, $F_{sim}$. Minimizing the objective function $F_{sim} - C_1 F_m$ with respect to $C_1$ yields the numerical constant of eq. \ref{eq:Fm_scaled}. Ultimately we want to use $F_{m,3D}$ in a balance with overburden forces to predict the depth at which particles can experience vertical lofting following a laterally propagating SW wavefront (discussed in sec. \ref{sec:SID}). As \cite{frizzell2023simulation} noted, the vertical component of the SW force imparted to particles in the wavefront is likely governed by the curvature of the wavefront in the near surface. This curvature can be described as the angle between the lateral direction and the wavefront normal direction ($\theta$). Preliminarily we have seen that this wavefront shape can be determined by multiplying the sound speed (eq. \ref{eq:sound_speed_nesterenko}) by a characteristic time scale, which is the number of particles in the wavefront multiplied by the Hertz collision time (eq. \ref{eq:contact_time_p2}). We did not implement any methodology for systematically determining the number of particles in the wavefront of our SW and calculating the collision time introduces further complications since we do not yet know of a method for determining the individual particle velocities within the 3D front (done for a 1D chain in \cite{nesterenko2013dynamics}) nor how to account for faster particle velocities than expected near the surface (an open topic in the literature \cite{tell2020acoustic}). As such, we will continue to assume $\theta$ is small ($\sim$1$^{\circ}$) as in \cite{frizzell2023simulation}.

\subsection{Data analysis} \label{sec:data_analysis}

As with sec. \ref{sec:SimSetup_p2}, we use largely the same process as detailed in \cite{frizzell2023simulation} to analyze wave speeds and induced volume change. We briefly review these procedures here. The particle channel is divided into grids and the states of particles falling within a grid bound (force, velocity, stress, $\phi_0$) are averaged to produce a `virtual sensor' state reading at each time step. Taking the average over all the depths at a given radial location in the channel produces the average state at that radial location. We then take the average over a radial measurement range to generate a single average value that can be used to compare results across the parameter space. Error bars that are shown when we report these averages are an indication of the spread of the value (either over the channel length or over the depth we consider) as opposed to a statistical measure.  We track the wave by finding the peak lateral force that occurs within each sensor. We determine wave speed by finding the slope of the position vs time data for the peak forces in the BL zone (for the initial wave speed, $c_w$) and in the middle of the channel (DSW propagating speed, $c_p$). Force, $\delta_m$, $v_m$, height change, and depth vs density variation are averaged over a measurement region which is roughly half the channel, with the midpoint located at the center of the channel. We have made a few modifications within this framework and we detail those here. Along with the \textit{MATLAB} generated plots we show in this paper, we have generated videos of selected simulations using \textit{OVITO}, \cite{stukowski2009visualization} (see the Supplementary information at the end of this paper for details).

\subsubsection{Packing fraction} \label{sec:packing_fraction}
\begin{figure}[h]
\begin{center}
\includegraphics[scale = .8]{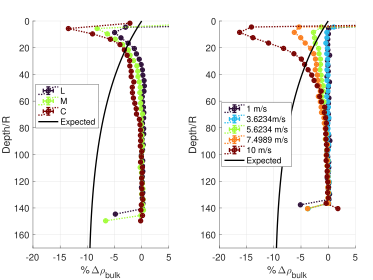}
\end{center}
\caption{\textbf{Bulk volume change vs depth.} Percent change in volume (\%$\Delta\rho$) on the x axis vs normalized (by particle radius) depth for varied levels of $\phi_0$.  \textbf{Left)} Channels of 2 meters in length filled 20 cm deep with particles were filled to the loose (L), medium (M), and compact (C) $\phi_0$ as described in sec. \ref{sec:prefilled} were subjected to piston impacts at $v_p$ = 10 m/s. The color scale represents $\phi_0$. As $\phi_0$ increases, dilation (negative values of \%$\Delta\rho$) increases in both the depth at which it begins (greater than $z/R$ = 100 for $\phi_0$ = C) and the peak dilated magnitude (around 14\% at $z/R$$\sim$10) affected. \textbf{Right)} The piston impact speed is varied in a single channel of particles ($R$ = 0.5 mm, $\phi_0$ = M), with the color legend corresponding to $v_p$. Dilation increases with increasing $v_p$. In both plots, the solid line represents the expected LCS percent volume change from Fig. \ref{fig:bandfield_density_pdiff}, normalized by an average lunar surface particle radius of $R$ = 250 $\mu$m (\cite{mckay1991lunar}).}
\label{fig:dilation_vs_depth_phi}
\end{figure}

Previously in \cite{frizzell2023simulation}, dilation was measured via channel height change. Here, we also find the packing fraction at depth in order to compare to the expected LCS density change of Fig. \ref{fig:bandfield_density_pdiff}. To do so, we perform a Voronoi tessellation using \textit{Voro++} (\cite{osti_946741}) of the channel as was performed in \cite{zhang2018rotational}. \textit{Voro++} computes a local packing fraction for each individual grain which becomes another averaged property of the virtual sensors. Tables \ref{tab:E_dt_check}, \ref{tab:rho_dt_check} and \ref{tab:particle_size_timestep} include initial packing fraction computed this way for the different channels we consider. Following the passage of the driven SW, we compute the percent change in the channel's bulk volume at each depth as a percent change over $\phi_0$ (shown in Figure \ref{fig:dilation_vs_depth_phi}). Fig. \ref{fig:dilation_vs_depth_phi} shows that the bulk density change vs depth agrees with the trends identified in \cite{frizzell2023simulation}, that SID increases with increasing $\phi_0$ and $v_0$. The magnitude of SID seen for these particles is smaller than the expected LCS dilation, which is expected since the material properties of the grains in these tests represent much softer particles than lunar regolith grains.

\subsubsection{Blast loading identification} \label{sec:blast_loading_method}
The blast loading region (lateral extent of the channel over which the PSW is active) should be a function of particle number \cite{nesterenko2013dynamics}. \cite{hasan2018shock} found the length of the decay region in a 1D granular chain to vary slightly with impactor momentum, with a range of 5 to 10 particle diameters for the momentum range considered. In \cite{frizzell2023simulation} the BL region was approximately 30 cm in length which is approximately 120 particle diameters with their particles ($R$ = 1.25 mm). Since we assume that we can scale the channel length by the same factor that we modify particle size (sec. \ref{sec:variedR_fill}), we verify that our channels extend a sufficient length beyond the BL region to allow for DSW measurement. For example, the channel in Table \ref{tab:particle_size_timestep} with micron sized particles results in a 20 cm channel, which is less than the approximate BL region in \cite{frizzell2023simulation} of about 30 cm. If the BL region is dependent only on particle number, then the shorter channel is acceptable, but if it is not, then channel length should be increased. In this paper we will measure the extent of the BL region by comparing the maximum overlap induced by the SW wavefront ($\delta_m$) to $\delta_0$. When the PSW is active, $\delta_m/\delta_0$ should fall rapidly along the length of the channel until reaching a steady value (when DSW becomes dominant). In this paper, we will define the end of the BL zone as the first post-peak occurrence of $\delta_m/\delta_0$ falling within three deviations of the DSW wavefront mean. This process is visualized later in sec. \ref{sec:blast_loading_results}. 

\subsection{Data generation and availability}
Following \cite{frizzell2023simulation}, channel preparation occurs on our standalone 44-core workstation and we run piston impact simulations using the University of Maryland High Performance computing cluster, Zaratan. Since we use a fixed channel geometry in this work, the number of particles in each simulation is roughly the same (around 600,000 particles) with some variation for $\phi_0$. Output file resolution varies drastically with the selected material parameters and observed phenomenon of interest (PSW vs DSW vs ejected ballistic particles) leading to output files anywhere from 30 to 100+ GB in total size and simulation duration between 45 minutes and three days. Since each parameter sweep involves upwards of 5 tests and we conducted around 30 different sweeps, we do not attempt to publicly host these data. Instead, we provide \textit{LIGGGHTS} input files that can run our simulations (\cite{frizzell2023code}) along with the restart files used to run them at (\cite{frizzell2023data}). Any reader could download these files recreate our results using the latest version of \textit{LIGGGHTS}. While it is not feasible to host all data required to perform the force and overlap tracking of the wavefront, the data required to recreate the results of sec. \ref{sec:dilation_results} is much smaller, requiring only the initial and final output files. The output files from these tests are provided along with the restart files at (\cite{frizzell2023data}).

Readers should be aware that the data used in our analysis is not output at the same rate as the simulation time step rate. In general, we followed the same output guidance as given in \cite{frizzell2023simulation} and scaled the output rate using the relevant power law scaling from eq. \ref{eq:sound_speed_nesterenko} when material parameters are changed. For example, increasing $E$ by 1000 would be a 10x increase in $c_0$ (since $c_0\sim E^{1/3}$) requiring a 10x increase in output rate (and corresponding reduction in total simulation time) compared to \cite{frizzell2023simulation} to capture the SW with sufficient resolution to identify the peak force in each virtual sensor (sec. \ref{sec:data_analysis}). However to allow for particles with large ballistic velocities (the hardest particles) to return to the channel surface and also to observer slower phenomenon that we originally expected (sec. \ref{sec:floor_vortex}), we performed many instances of output adjustment, manually extending the simulation run time to allow more time for particles to reach the channel or increasing the initial high resolution output rate when our peak force detection algorithm required finer resolution. Readers can refer to the input files at (\cite{frizzell2023code}) and review the input file corresponding to the case in question to determine specific details about data output rates, if needed.

\subsection{Investigated parameters} \label{sec:investigated_params}
SID requires a low gravity and vacuum environment. While we simulated SID under these conditions, the material properties of the grains in \cite{frizzell2023simulation} were not representative of the lunar environment (those particles were an order of magnitude too large and had an elastic modulus closer to rubber than to regolith). We first endeavor to understand how SID will scale by characterizing the forces experienced by particles in our simulated SW wavefronts with respect to order of magnitude changes in density, mass, size, and impact velocity. We then use the results of these simulations to find the coefficient in eqn. \ref{eq:Fm_scaled} (discussed in sec. \ref{sec:wavefront_forces}). We show the ranges of $E$, $\rho$, $v_0$, and $R$ we consider in this analysis in Table \ref{tab:investigated_params}. 

\begin{table*}[h!] 
\begin{minipage}{425pt}
\caption{Investigated parameters. We give the range of parameter values for the physical quantities evaluated in our SID sensitivity analysis. We give the range of values used as input values in our simulations along with the relevant characteristic condition (with lunar particles referenced for material properties and similar low gravity DEM simulations referenced for simulated parameters). If there is a relevant reference we provide this here as well. Grain properties are given for lunar surface particles (regolith at depths of 0 to 40 cm).}\label{tab:investigated_params} \small
\begin{tabular}{@{}cccccc@{} }
\toprule
Quantity & Symbol (units) & Range investigated & Characteristic &  References \\
\midrule
Elastic modulus & $E$ (MPa) & 5 - 5000 & 500 & \cite{frizzell2023simulation}, \cite{cole2012particle}   \\
Particle density & $\rho$ (g/cc) & 0.8 - 25 & 3   & \cite{carrier1991physical}, \cite{colwell2009lunar}, \cite{hanus2017volumes}   \\
Particle radius & $R$ (mm) & 0.125 - 25 & 0.100  &  \cite{mckay1991lunar}  \\
Piston velocity & $v_0$ (m/s) & 1 - 360  & n/a & n/a   \\
\hline

Cohesion & $k_c$ (kPa) & 0 - 10 & 0.44 - 1.1 & \cite{carrier1991physical}, Table 9.12  \\
Friction & $\mu$ & 0 - 1.8  & 0.87 - 1.28 & \cite{yu2014numerical}    \\
Poison ratio & $\nu$ & 0.15 - 0.45 & 0.2 - 0.4   & \cite{kovach1973velocity}, \cite{sutton1970elastic}, \cite{morgan2018pre}  \\
Gravity & $g$ (m/s$^2$) & 0.1 - 9.8 & 1.625  &    \\
Dispersity & - & Mono-, bi-, tri- disperse & log-normal & \cite{mckay1991lunar}    \\
Rolling Friction & $\mu_r$ & 0 - 1.0 & 0.8   & \cite{sanchez2016disruption}  \\
Rolling Damping & $\gamma_{d,r}$ & 0 - 2.5 & 2  &  \cite{frizzell2023simulation}, \cite{holmes2016bending} \\
Restitution & $e$ & 0.35 - 0.85 & 0.5  & \cite{chau2002coefficient}, \cite{zhang2018rotational}, \cite{wang2020behaviors}   \\
Momentum & $\rho_i/\rho$ & 0.01 - 10 &  n/a & n/a   \\
\bottomrule
\end{tabular}
\end{minipage}
\end{table*}

Our particle density range considers the softest and hardest (smallest/largest elastic modulus, respectively) particles found in low gravity settings throughout the solar system. The elastic modulus range encompasses the soft particles of \cite{frizzell2023simulation} and approaches the hardness of lunar minerals ($E$ $\sim$ 50 GPa, for lunar plagioclase \cite{cole2012particle}). Particle sizes extend two orders of magnitude above the typical lunar particle size discussed in sec. \ref{sec:variedR_fill} ($R$ = 125 $\mu$m to 12.5 mm). In general we used $v_0$ between 1 and 10 m/s as in \cite{frizzell2023simulation}, though  particles with larger $E$ can undergo collisions at greater speeds since they have a larger $c_m$. We also conducted a simulation campaign in which both $E$ and $v_0$ are increased to maintain $v_0$ at the same portion of $c_m$ (discussed in sec. \ref{sec:model}). We next conducted a sweep in $E$ space using channels of higher and lower density particles ($\rho$ = 5, 1.5 g/cm$^3$) as well as velocity sweeps in channels filled with particles at the extremes of densities we consider ($\rho$ = 0.8, 25 g/cm$^3$). The last assembly we prepared was one filled with lunar-like particles ($R$ = 125 $\mu$m, $\rho$ = 3 g/cm$^3$, $E$ = 500 MPa) which was prepared using the method of sec. \ref{sec:prefilled}. Note that this channel requires a time step of $10^{-8}$ s which is less than 0.3\% of the Hertz collision time. Knowing how the wavefront force changes with these four parameters, we can then evaluate how the force in the SWs compare to the overburden force (described in sec. \ref{sec:SID}) in order to understand how the depth of the balance point changes in different environments (e.g., channels with particles of varied material properties or different gravitational loadings, see sec. \ref{sec:scaling_predictions}). We also consider one case of varying impactor momentum in which we vary the mass ratio between the virtual piston and channel grains (designated as $\rho_i/\rho$, where $\rho_i$ is the density of the piston/impactor particles) analogous to the density ratio of impactor meteor and regolith.

Finally, we compare the predicted behavior of SID to that produced in our simulations (sec. \ref{sec:dilation_results}). In addition to the parameters in eq. \ref{eq:Fm_scaled}, we vary several other important physical parameters which are likely to influence SID (also summarized in Table \ref{tab:investigated_params}). We consider how changes to the Poisson ratio, cohesion, friction, and gravity affect surface bulk dilation. While the lunar regolith size distribution is essentially a continuum, we have so far only considered finite size distributions. We evaluate how SID changes as we move away from a monodisperse assembly with tests conducted in channels filled with bi- and tri-disperse assemblies. Lastly, we also considered DEM simulation parameters (typically used to allow spherical particles to mimic real particle behavior) that are likely to influence simulated SID (rolling friction, rolling viscous damping and coefficient of restitution).

\section{Results and Discussion}  \label{sec:Results_p2}

We review the results of our simulations and the outcome of fitting simulated forces to $F_{m,3D}$. We find that wavefront forces in the 3D channel scale the same way as forces in a 1D chain and SID trends are as expected (more SID for harder and smaller particles). We then propose a method for predicting SID based on $F_{m,3D}$ and compare predicted and simulated dilation. Both dilations scale the same way and are within an order of magnitude. Finally, we briefly revisit the outbursts capable of affecting SID after experiencing lower $v_0$ impacts.

\subsection{3D SW wavefront forces} \label{sec:wavefront_force_fitting}
\begin{figure*}[h!]
\begin{center}
\includegraphics[scale = 0.8]{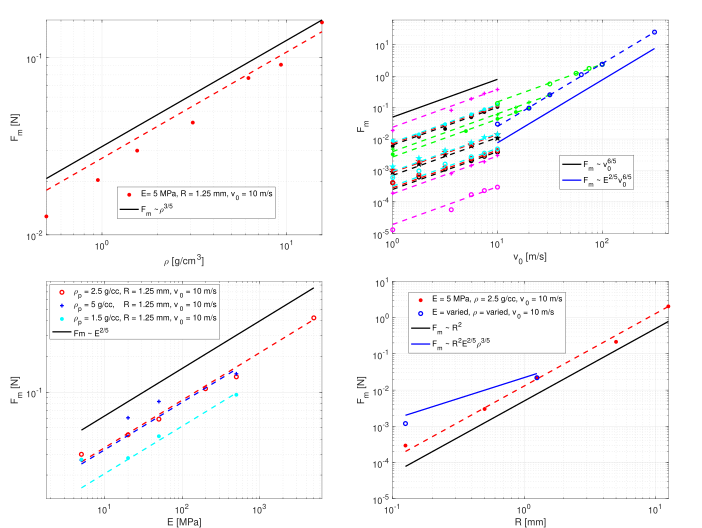}
\end{center}
\caption{\textbf{$F_m$ vs $\rho$, $v_0$, $E$, and $R$ for varied cases.}  Each plot shows average force in our 3D simulated waves vs the parameter of interest ($v_0$ in the upper right (UR), $\rho$ in the upper left (UL), $E$ in the lower left (LL) and $R$ in the lower right (LR). The average simulated force for the given test is represented by a single data point and dashed lines are the corresponding fit. Solid lines show expected power law trends and are described in the legend of each individual plot. \textbf{UR)} Legend entries for the $v_0$ case are turned off due to over crowding. Star markers represent base case material properties, with cyan, red, and black colors corresponding to compact, medium, and loose $\phi_0$, respectively. Solid markers are increased density ($\rho$ = 25 g/cc) and open circles are reduced density ($\rho$ = 0.8 g/cc) which (respectively) increase and decrease the force experienced by particles as a result of the wavefront. The pink trends correspond to varied particle sizes, where the lowest forces are experienced by the smallest particles ($R$ = 125 $\mu$m, pink circles) and the highest forces experienced by the largest particles ($R$ = 125 $\mu$m, pink plus sign). The green trends represent increasing modulus of elasticity, where E ranges from 5 (green filled circle) to 500 MPa (open green circle). The data and trend in blue represent a campaign of varied $v_0$ and $E$, with $v_0$ scaled to remain a constant portion of $c_m$.  \textbf{LL)} We show the base case (with varied $R$) in red, with blue and cyan corresponding to slightly greater and smaller densities (respectively). \textbf{LR)} We included a comparison case of varied $E$ and $\rho$ in addition to $R$ as the blue points which show how the force changes between the base particles and the lunar-like particles ($\phi_0\sim$58\% and $v_0$ = 10 m/s. The upper solid blue line shows the expected scaling based on changing $E$, $\rho$ and $R$ while the lower solid black line shows the expected $R^2$ scaling. We did not calculate a fit for the comparison case given the paucity of points, but it is clear that the simulated results match the expected scaling.}
\label{fig:fm_vs_trends}
\end{figure*}

First, we assess our hypothesis that the forces in a 3D DSW wavefront follow the same scaling as those experienced by particles experiencing a 1D wavefront. We conducted parameter sweeps for each of the four variables in eq. \ref{eq:Fm_scaled} ($E$, $R$, $\rho$ and $v_0$) while holding all other parameters constant. The parameter sweeps over $E$, $\rho$ and $v_0$ use pre-filled channels while the particle size sweep ($R$) required newly filled channels. Once we obtain a single value for the magnitude of the force in the wavefront for each set of parameters within a campaign (as discussed in sec. \ref{sec:data_analysis}), we perform a least squares fit to eq. \ref{eq:Fm_scaled} (see sec. \ref{sec:wavefront_forces}). The results are presented for each parameter in Fig. \ref{fig:fm_vs_trends} which shows excellent agreement with expected power law force scaling.

First we consider the upper right plot of Fig. \ref{fig:fm_vs_trends} which shows $F_m$ vs $v_0$. The majority of the cases we evaluated in this paper are velocity sweeps in channels with varied material parameters since they were the easiest to setup and quickest to run (see sec. \ref{sec:prefilled}). We first created DSW using varied $v_0$ in channels filled with particles with base parameters filled to loose, medium, and compact $\phi_0$ (sec. \ref{sec:prefilled}). We then repeated this procedure, conducting sweeps through $v_0$ using channels with the largest ($\rho$ = 25 g/cm$^3$) and smallest ($\rho$ = 0.8 g/cm$^3$) densities filled to the same three $\phi_0$. Increasing packing fraction increases $F_m$ slightly (particles are closer together, have more neighbors, and thus experience greater forces as the wavefront passes by) while the increase or decrease in density leads to order of magnitude shifts in $F_m$. $F_m$ increases as $\rho$ increases since the impactor particle has more energy which must be absorbed by the spring potential as the DSW wavefront passes. As we continue velocity sweeps, we see that while increasing $E$  the particles are `harder' and therefore, for the same $\delta_m$, the force imparted to the particles by the wavefront is larger (see eq. \ref{eq:force_model}). All the aforementioned parameter sweeps in $v_0$ agree with expectation based on eq. \ref{eq:Fm_scaled}, with the force scaling with velocity as $F_m \sim v_0^{6/5}$. Including a second scaling parameter in the form of varied $v_0$ and $E$ (when the value of impact velocity is scaled to be a constant fraction of $c_m$, even with increasing $E$) also agrees with expectation, reproducing nicely the power law $F_m \sim v_0^{6/5}E^{2/5}$ from eq. \ref{eq:Fm_scaled}. The good agreement with $v_0$ and $E$ $+$ $v_0$ scaling over varied values of $R$ and $\rho$ justifies our decision to conduct fewer cases when conducting parameter sweeps in $\rho$, $R$, and $E$.

\begin{figure}[h]
\begin{center}
\includegraphics[scale = 0.4]{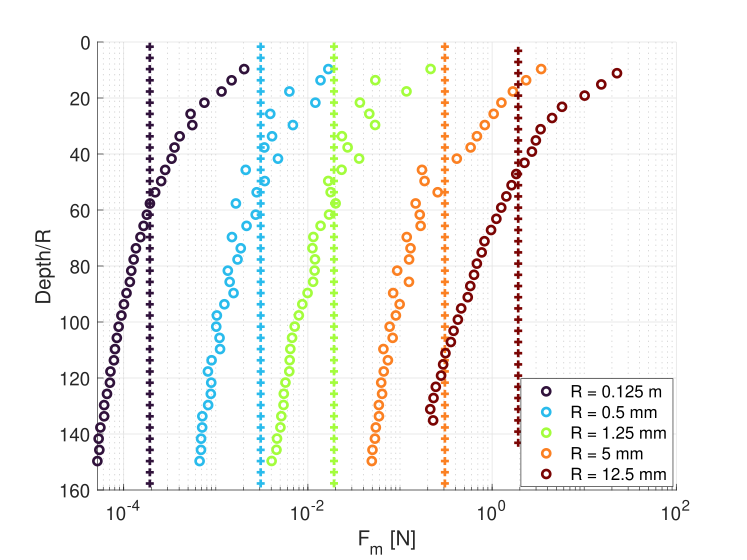}
\end{center}
\caption{\textbf{Normalized depth ($z/R$) vs average $F_m$.} We show average simulated force at each depth for 3D assemblies of varied size, where the color legend corresponds to particle diameter. Channels are filled to the medium $\phi_0$ level of packing and are listed in Table \ref{tab:particle_size_timestep}. Simulated forces are circles and the plus sign markers show the computed force via eq. \ref{eq:our_FSW} for the given particle radius, base material parameters, and in input velocity of $v_0$ = 10 m/s.  }
\label{fig:fm_vs_depth_R}
\end{figure}

The parameters besides $v_0$ also show good agreement with expectation and we now walk through the remainder of Fig. \ref{fig:fm_vs_trends}. In addition to the base case parameters where $\rho$ = 2.5 g/cm$^3$, we performed $E$ sweeps in channels of slightly lower ($\rho$ = 1.5 g/cm$^3$) and higher ($\rho$ = 5 g/cm$^3$) density (bottom left plot). In each case the expected $F_m \sim E^{2/5}$ power law is roughly recreated, with $F_m$ increasing with increasing density. The fit is not quite as good in these cases ($r^2$ of 0.986 and 0.945 for lower and higher $\rho$, respectively) as in the $v_0$ parameter sweeps, but this is a result of difficulty in tracking the wavefront location near the surface, and we discuss this in Fig. \ref{fig:fm_vs_depth_R}. Fig. \ref{fig:fm_vs_depth_R} shows the average force vs depth on a particle in the wavefront for particles of varied size. A single data point in any of the plots in Fig. \ref{fig:fm_vs_trends} corresponds to the average force over all depths corresponding to a single trend line in Fig. \ref{fig:fm_vs_depth_R} (for example, $F_m$ for $R$ = 1.25 mm particles is an average over the green points). In some cases our method of tracking the wavefront by identifying the first peak in force that exceeds a threshold value misidentifies that actual peak for an earlier point, resulting in reduced forces across the sensors of a given depth. This most often happens near the surface (cyan and green) but can sometimes happen at middle depths too (orange). The noise acts to lower the average value of $F_m$ reported and this issue is most evident in the $E$ sweep case with varied $\rho$. The noise in the results could be resolved by designing a tracking method that applies a filter to the data to help reduce misidentified wavefront location, but we leave that as an item for future work. Despite the noise leading to the outliers in the $F_m$ vs $E$ plot of Fig. \ref{fig:fm_vs_trends}, the fit with eq. \ref{eq:Fm_scaled} is still quite good ($r^2>$ 0.94, tab. \ref{tab:goodness_of_fit}) and aligns with the expected power law, $F_m \sim E^{2/5}$. 

Assuming that agreement with expectation would extend to the $\rho$ and $R$ sweeps, we performed only a single campaign of tests in each channel. Fig. \ref{fig:fm_vs_trends} shows that we still follow the expected scaling: the upper left plot for $F_m$ vs $\rho$ follows $F_m\sim\rho^{3/5}$ and the bottom right plot for $F_m$ vs $R$ follows $F_m\sim R^{2}$. We included one final case that compares the lunar-like particles to the base particles in the $F_m$ vs $R$. Though we do not construct a fit, the two points demonstrate that varying $E$, $\rho$, and $R$ together still produces the expected scaling. Having confirmed visually that the power law scaling we expected is produced, we assemble all results for the fit to parameter $C_1$ in eq. \ref{eq:Fm_scaled} as well as the goodness of fit (assessed by calculating the $r^2$ correlation coefficient of the fit) in Table \ref{tab:goodness_of_fit}. 

\begin{table}[h]
\begin{minipage}{210pt} 
\caption{Goodness of fit. Scaling parameter and $r^2$ for the cases inspected. Particles are set using the base parameters and then one or more of the components are varied corresponding to the values in fig. \ref{fig:fm_vs_trends}. For the case with varied $v_0$ and varied $E$, $E$ values are found in Table \ref{tab:E_dt_check}. Packing fraction corresponds to compact (C), medium (M), and loose (L) $\phi_0$.} \label{tab:computation_details} 
\begin{tabular}{@{}ccccc|cc@{}}
\toprule
\begin{tabular}{@{}c@{}} E \\ (MPa) \end{tabular} &
\begin{tabular}{@{}c@{}} $\rho$ \\ (g/cc) \end{tabular} &
\begin{tabular}{@{}c@{}} R \\ (mm) \end{tabular} &
\begin{tabular}{@{}c@{}} $v_0$ \\ (m/s) \end{tabular} &
\begin{tabular}{@{}c@{}} $\phi_0$ \end{tabular} &
\begin{tabular}{@{}c@{}} $C_1$ \\ $\times10^3$ \end{tabular}& 
\begin{tabular}{@{}c@{}} $r^2$ \\ (\%) \end{tabular}   \\
\midrule
 varied & 1.5 & 1.25 & 10 & C & 8.1 & 98.6   \\
 varied & 2.5 & 1.25 & 10 & C & 9.8 & 99.2 \\
 varied  & 5  & 1.25 & 10 & C & 6.2 & 94.5  \\ 
 \hline
 5  & 2.5  & varied & 10 & M & 7.7 & 99.7  \\ 
 \hline
 5 & varied  & 1.25 & 10 & C & 17.8 & 97.6   \\
 \hline
 5 & 2.5  & 1.25 & varied & L & 4.2 & 99.5   \\
 5 & 2.5  & 1.25 & varied & M & 5.3 &  99.9  \\
 5 & 2.5  & 1.25 & varied & C & 5.7 & 99.4   \\
 5 & 25  & 1.25 & varied & L & 9.8 &  99.8  \\
 5 & 25  & 1.25 & varied & M & 10.9 & 99.9   \\
 5 & 25  & 1.25 & varied & C & 11.8 & 99.9   \\
 5 & 0.8  & 1.25 & varied & L & 3.0 & 99.9   \\
 5 & 0.8  & 1.25 & varied & M & 3.3 & 99.8   \\
 5 & 0.8  & 1.25 & varied & C & 3.8 & 99.9   \\
 varied & 2.5  & 1.25 & varied & C & 9.6 & 99.9   \\
 5 & 2.5  & 10 & varied & M & 9.0 &  99.1  \\
 5 & 2.5  & 1 & varied & M & 1.7 & 99.3   \\
 5 & 2.5  & 0.125 & varied & M & 11.5 &  97.1  \\
 20 & 2.5  & 1.25 & varied & C & 9.3 & 99.8   \\
 50 & 2.5  & 1.25 & varied & C & 9.6 & 99.9   \\
 500 & 2.5  & 1.25 & varied & C & 9.2 & 99.8   \\
 
\bottomrule
\end{tabular}
\label{tab:goodness_of_fit}
\end{minipage}
\end{table}

The values found for $C_1$ are of approximately the same magnitude and fall in the range of 0.002 to 0.012. While the noise in some of the our data likely adds to the variability in $C_1$, it is also likely that the scaling coefficient itself is dependent on the particle properties of the assembly. This makes sense; adjusting $E$, $\rho$, $R$ influences the final geometry of the random packing (as particles become harder and softer they settle into slightly different positions) and it is the specific geometry of a 3D  assembly that leads to nonlinear force propagation (\cite{goddard1990nonlinear}). From this work it is not possible to determine the dependence of $C_1$ on material parameters and impact velocity. However, determining the exact relationship of $C_1$ to our four parameters of interest is outside the scope of this work - here we seek to determine only if the scaling of $F_m$ in a 3D assembly is similar to that in a 1D assembly. Indeed, we have shown that $F_m$ follows (to a numerical constant) the same scaling as in a 1D chain (eq. \ref{eq:Pal_FSW}). For the remainder of this paper we will use eq. \ref{eq:our_FSW} as the forces within a 3D DSW wavefront to assess how SID will scale to granular assemblies with different properties in different environments. We take $C_1$ to be the mean of the fit values from Table \ref{tab:goodness_of_fit} with $r^2>0.99$ which results in mean and deviation of $C_1$ of about 7.3x10$^{-3}$ and 3.2x10$^{-3}$.

\begin{align} \label{eq:our_FSW}
    F_{m} = C_1 k\left( \frac{5}{4}mv_0^2 k^{-1} \right)^{3/5} \\
    C_1 = (7.3 \pm 3.2)\times10^{-3} \nonumber
\end{align}

While we do proceed using eq. \ref{eq:our_FSW}, Fig. \ref{fig:fm_vs_depth_R} highlights that the result of our efforts in this paper are a prediction of the \textit{average} peak force a particle experiences within the wavefront of a DSW in a 3D randomly packed \textit{shallow} channel. Specifically, in a channel that is approximately 80 particle diameters tall. In Fig. \ref{fig:fm_vs_depth_R} we plot the predicted force at each depth using eq. \ref{eq:our_FSW} alongside the simulated values. The predicted force from eq. \ref{eq:our_FSW} does not vary with depth while the simulated forces are higher near the surface but decrease towards the floor. Particles experiencing wave propagation near the grain-vacuum interface are known to behave anomalously (mentioned in sec. \ref{sec:wavefront_forces}) and our method of fitting to the averaged force is influenced by the higher than expected magnitudes at the surface. Our fit parameter $C_1$ therefore will lead eq. \ref{eq:our_FSW} to underpredict $F_m$ in shorter channels and over predict $F_m$ in taller channels. Nevertheless, the uniformity of the scaling is evident from fig. \ref{fig:fm_vs_trends} and eq. \ref{eq:our_FSW} can be used in an initial characterization of the scalability of SID.

\begin{figure*}[h]
\begin{center}
\includegraphics[scale = .7]{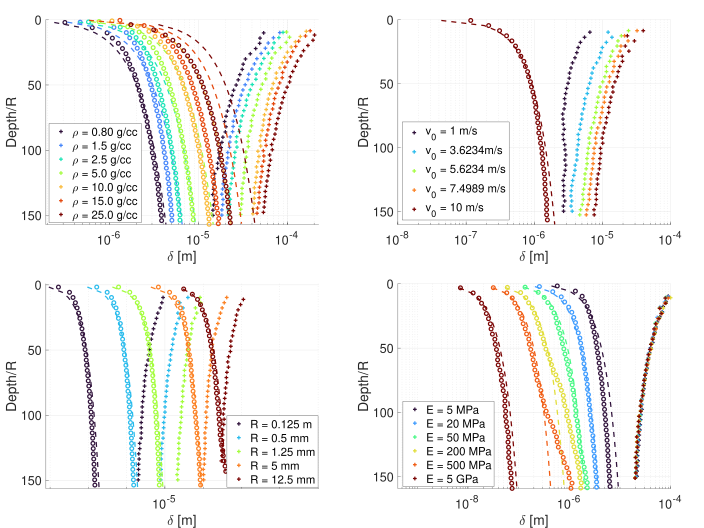}
\end{center}
\caption{\textbf{Normalized depth vs $\delta$ for varied $\rho$, $v_0$, $E$, and $R$.} Depth $z$ on the y axis is normalized by particle radius ($R$) with overlap ($\delta$) on the x axis. We show the trends for varied $v_0$ in the upper right (UR), $\rho$ in the upper left (UL), $E$ in the lower left (LL) and $R$ in the lower right (LR). Each color legend gives the value to the respective varied parameter with circles, pluses, and dashes corresponding to $\delta_0$, $\delta_m$, and the prediction from eq. \ref{eq:overburden} (respectively). $\delta_m$ and $\delta_0$ `pinch' together as depth increases and SID only occurs when there is a substantial difference between the two.} 
\label{fig:dm_and_d0_vs_depth}
\end{figure*}

Before turning to a scaling analysis, we want emphasize that the overlap experienced by particles in the wavefront ($\delta_m$) is dominant in determining the extent of SID. Figure \ref{fig:dm_and_d0_vs_depth} mirrors Fig. \ref{fig:fm_vs_trends}, with each plot showing the initial and maximum overlap experienced in the channel as the result of varying a single parameter ($E$, $R$, $\rho$, and $v_0$). The maximum force in the channel from Fig. \ref{fig:fm_vs_depth_R} echoes $\delta_m$, with larger $\delta_m$ in the wavefront near the surface than near the floor. As we will see in the following sections, the cases that exhibit the largest magnitude of SID correspond to those cases in Fig. \ref{fig:dm_and_d0_vs_depth} that have the largest difference between $\delta_0$ and $\delta_m$ (i.e., large $E$ and $v_0$, smaller $\rho$ and $R$), with particle size and elastic modulus having the largest influence. Notably, we show that all cases of varied $E$ with $v_0$ held to a constant portion of $c_m$ produces the same $\delta_m$. Particles with larger $E$ have smaller $\delta_0$ (they are `harder') and therefore the largest difference between $\delta_0$ and $\delta_m$ and exhibit the largest SID response. Fig. \ref{fig:dm_and_d0_vs_depth} also highlights a few instances of the non-physical initial states we mentioned earlier in sec. \ref{sec:prefilled}. The dashed lines in the plots represent the analytical prediction of $\delta_0$ using eq. \ref{eq:delta0} and our simulated channels agree with this prediction nicely for the majority of cases. However, in the particle channels with varied density, the distributions corresponding to the two largest $\rho$ (orange and dark red) have lower $\phi_0$ than would be expected, but this is a result of filling the channel with one set of properties and then adjusting the particle's material properties after a static packing is achieved. It is possible that this kind of discrepancy is responsible for some of the variability in $C_1$, but we assume in this case that any influence of a `too loose' or `too compact' channel is mitigated by our averaging procedure. That assumption can be reevaluated by others; here we will just keep this in mind when considering resultant SID.

Finally, we briefly confirm the validity of our results by considering produced wave speeds, maximum impact velocities of particles within a DSW wavefront, and the total overlap experienced by particles during a simulation. We provide the results of this assessment in Table \ref{tab:validity}. The second column grouping compares computed initial PSW and DSW speeds to the calculated wave speed. As we can see, $c_0$ increases with $E$, $\mu$ and $g$, slightly increases with $R$ given our fixed channel height (discussed in sec. \ref{sec:model}), but decreases with $\rho$ as expected.  The third grouping of columns considers the maximum collision speed of particles in the wavefront compared to $L_e$, the speed above which the Hertz law can become invalid for describing the contact. In all cases we exceed $L_e$, but the problem is the worst with low $E$ or large $\rho$. As we move towards particles that look more like lunar surface grains, $v_m$ is only a few times larger than $L_e$. The last grouping of columns compares the maximum overlap which should not exceed 1\%$R$. Similar to the impact velocity considerations, the cases that are closest to (or exceed) the limit are those with the softest particles. Again, as particles become more lunar-like (larger $E$), $\delta_m$ is a smaller portion of $R$ and the lunar-like particles are less than 0.3\% of $R$. As we saw in Table \ref{tab:youngs}, as $E$ increases, $\phi_0$ becomes smaller since grains inter-penetrate less (thanks to larger $k$) which means that, for the same velocity, a channel filled with larger $E$ particles experiences smaller $\delta_m$ in the wavefront. We are confident that our simulations model the real effects of impacts propagating along the floor at a fixed speed given the adherence of $F_{m,3D}$ to expected power law trends as well as the validation performed in \cite{frizzell2023simulation}. That $v_m$ does slightly exceed $L_e$ indicates that particles undergo high stresses which may lead to some fracturing, though we believe this would generate microporosity that is insignificant compared to the reduction in thermal inertia that is achievable through SID. Evaluating grain scale stresses vs fracture limits is an important consideration for future work. 

\begin{table*}[h!]  \small
\begin{minipage}{450pt} 
\caption{Validity details. We assess the physical realness of our results be comparing wave speed, particle speed within the wavefront ($v_m$) and the maximum overlap particles in the wavefront experience. In all cases the base material parameters are used and the varied parameter of interest is changed to the value given in the table for each test. Impact velocity is set to $v_0$ = 10 m/s in all cases, except for the velocity sweep which used based material parameters and the velocity listed. $\phi_0$ for each of the cases can be found in across Tables \ref{tab:youngs}, \ref{tab:particle_density_timestep} and \ref{tab:particle_size_timestep}, with most being filled to compact fillings except the particle size sweep and lunar-like particles which are filled to a medium $\phi_0$. The measured (simulated) values are taken as an average of a given quantity over the entire channel depth - we disregard the deviations in this table and consider the mean values to get a sense of the scaling. For each test we calculate sound speed $c_0$ (as the average over a depth of 160$R$ using eq. \ref{eq:sound_speed_nesterenko}) and compare it to the generated PSW wave speed ($c_w$) and the DSW speed ($c_p$) which are found using the same procedure as in \cite{frizzell2023simulation}. We note that $c_w$ and $c_p$ can fall below the $v_0$ impact speed since they average over the entire depth of the channel and $v_0$ is only maintained  by impacts along the floor. We do not give a $c_p$ for the $E$ = 5 GPa case as the measurement contained too much noise, caused by outbursts (sec. \ref{sec:floor_vortex}). We also collect $v_m$ and compare it to the elastic limit ($L_e$, also found in Tables \ref{tab:youngs}, \ref{tab:particle_density_timestep} and \ref{tab:particle_size_timestep}). Finally, we collect the maximum overlap experienced over the course of each test and compare it to particle radius. The case of $E$ = 50 GPa was included as a reference and was not simulated, hence the `n/a' for simulated quantities.} \label{tab:validity}
\begin{tabular}{@{}cc|ccc|ccc|ccc@{}}
\toprule
 & & \multicolumn{3}{c|}{Bulk speed} &  \multicolumn{3}{c|}{Particle speed} &  \multicolumn{3}{c}{Particle overlap} \\
\begin{tabular}{@{}c@{}} Param. \end{tabular} &
\begin{tabular}{@{}c@{}} Value (units) \end{tabular} &
\begin{tabular}{@{}c@{}} $c_0$ \\ (m/s) \end{tabular} &
\begin{tabular}{@{}c@{}} $c_w$ \\ (m/s) \end{tabular} &
\begin{tabular}{@{}c@{}} $c_p$ \\ (m/s) \end{tabular} &
\begin{tabular}{@{}c@{}} $v_m$ \\ (mm/s) \end{tabular} &
\begin{tabular}{@{}c@{}} $L_e$ \\ (mm/s) \end{tabular}  &
\begin{tabular}{@{}c@{}} $v_m$/$L_e$  \end{tabular}  &
\begin{tabular}{@{}c@{}} $\delta_m$ \\ ($\mu$m) \end{tabular}  &
\begin{tabular}{@{}c@{}} $\frac{\delta_m}{R}$ \\  (\%) \end{tabular}  & 
\begin{tabular}{@{}c@{}} $F_m$ \\ (mN) \end{tabular} \ \\

\midrule

 $E$ & 5 (MPa) & 8.65   & 14.00 & 10.84 & 181.2 & 9.43  & 19.2  &16.94 & 1.36 & 14.00  \\
 $E$ & 50 (MPa) & 18.64   & 30.42 & 22.93 & 213.08 & 29.81 & 7.1  & 6.57 & 0.53 & 59.20  \\
 $E$ & 500 (MPa) & 40.17   & 65.82 & 55.82 & 341.49 & 94.28 & 3.6  & 2.55 & 0.20  & 134.61  \\
 $E$ & 5 (GPa)  & 86.54  & 136.18 & - & 501.7 & 298.14 & 1.7   & 0.82 & 0.07  & 422.83  \\
  $E$ & 50 (GPa)  & 186.44   & n/a  & n/a & n/a & 942.81  & n/a & n/a & n/a  & n/a  \\
 \hline
 $\rho$ & 0.8 (g/cc) & 12.65  & 20.53  & 15.98 & 175.7  & 16.7 & 10.5 & 10.19 & 0.82 & 12.72     \\
 $\rho$ & 2.5 (g/cc) & 8.65   & 13.57 & 10.60  & 183.5 & 9.43 & 19.5 & 17.17 & 1.37 & 29.91   \\
 $\rho$ & 10 (g/cc) & 5.45   & 8.13  & 8.30  & 159.2  & 4.7 & 33.9 &  26.56 &  2.12 & 76.82  \\
 $\rho$ & 25 (g/cc) & 4.02   & 6.19  &  7.40  & 85.5  & 3.0 & 28.5 &  28.77 &  2.30 & 158.47    \\
 \hline
 $R$ & 125 ($\mu$m)  & 5.83   & 15.40  & 8.76  & 154.92  &  9.43  & 16.4 & 1.41  & 1.13   &  0.268    \\
 $R$ & 0.5 (mm)  & 7.35  & 15.22  & 9.04 & 110.60  & 9.43  & 11.7 & 4.62  &  0.92  & 2.73       \\
 $R$ & 1.25 (mm)  & 8.65   & 15.71 & 9.58  & 149.14  & 9.43   & 15.8 & 15.65  &  1.25   & 27.58       \\
 $R$ & 5 (mm)  & 10.78   & 14.75  & 9.87  & 118.05  & 9.43  & 12.5  & 80.20  & 1.60   & 375.69    \\
 $R$ & 12.5 (mm)  & 12.56   & 13.79  & 10.14  &  52.03  & 9.43  & 5.5  & 225.58  &  1.80  & 2,371.3     \\
 \hline
 $v_0$ & 1 (m/s) & 7.35   & 9.60 & 7.06 & 24.5  & 9.43 & 2.6   & 2.11  & 0.17  & 0.18  \\
 $v_0$ & 3.6 (m/s) & 7.35   & 11.03 & 7.41 &  29.15  & 9.43 & 3.1   & 2.04  & 0.16   & 0.75   \\
 $v_0$ & 5.6 (m/s) &  7.35  & 12.06  & 8.28 & 60.38 & 9.43 & 6.4  & 3.09  & 0.25   &  1.43    \\
  $v_0$ & 7.5 (m/s) & 7.35 & 12.91 & 8.64 & 91.52  & 9.43 & 9.7   & 4.08  & 0.33   &  1.89    \\
 $v_0$ & 10 (m/s) & 7.35  & 13.61 & 9.23  &  129.81  & 9.43 & 13.8  & 5.13  & 0.41  &  3.00  \\ 
  \hline
 $g$ & 0.1 (m/s$^2$) & 5.44   & 13.25 & 9.95 & 186.0 & 9.43 & 19.7  & 16.64  & 1.33 &  23.45 \\
 $g$ & 1.625 (m/s$^2$) & 8.65  & 14.00 & 10.84  & 181.2 & 9.43 & 19.2   &  16.94 & 1.36 &  14.00  \\
 $g$ & 3.721 (m/s$^2$) & 9.93   & 14.36 & 11.45  & 171.8 & 9.43 & 18.2  & 17.15 & 1.37 &  14.02  \\
 $g$ & 9.8 (m/s$^2$) & 11.68   & 15.18 &  12.34  & 152.3 &  9.43 & 16.2  & 19.08 & 1.53 &  14.51  \\ 
 \hline
 $k_c$ & 0 (kPa) & 8.65   & 14.01 & 10.85 & 180.72 & 9.43 &  19.2 & 16.85  & 1.35   & 14.00  \\
 $k_c$ & 0.1 (kPa) & 8.65  & 14.00 & 10.86  & 180.63  & 9.43 & 19.2   & 16.85  & 1.35   & 14.00  \\
 $k_c$ & 1 (kPa) & 8.65 & 14.00 & 10.84  & 181.17  & 9.43 & 19.2   & 16.94  & 1.36   &  14.00   \\
 $k_c$ & 10 (kPa) & 8.65   & 13.84 & 10.82  & 183.88  & 9.43 & 19.5  & 17.72  & 1.42  &  14.12  \\
  \hline 
 $\mu$ & 0  & 8.65   & 10.83 & 9.13  & 249.49  & 9.43 & 26.46   & 24.81  & 1.98   & 17.66   \\
 $\mu$ & 0.4  & 8.65  & 11.57 & 9.49 & 159.31 & 9.43  & 16.89 & 15.89 & 1.27  & 11.58  \\
 $\mu$ & 1  & 8.65   & 14.00 & 10.84 & 181.2 & 9.43  & 19.22 & 16.94 & 1.36  & 14.00 \\
 $\mu$ & 1.8  & 8.65   & 15.03 & 11.70 & 211.0 & 9.43 & 22.38  & 18.96 & 1.52   & 16.94   \\
 \hline
 \hline
 $E$,$\rho$,$R$ & Lunar-like & 25.47 & 81.10  & 38.9 & 380.43 & 86.07 & 4.4 & 0.34 & 0.27 & 1.19\\
 
\bottomrule
\end{tabular}
\end{minipage}
\end{table*}

\subsection{Scaling} \label{sec:scaling_predictions}

Now that we have determined an expression for $F_m$ in a 3D DSW wavefront (eq. \ref{eq:our_FSW}), we can compare the force particles experience as a result of the wave to that of the gravitational overburden. SID occurs when particles at depth $z$ experience a vertical force from a SW wavefront that exceeds that overburden force (eq. \ref{eq:overburden}) at that same depth (net upward force leads to lofting, with particles setting down to a looser state after a ballistic trajectory \cite{frizzell2023simulation}). We posit that  particles experience this upward force thanks to the near surface curvature of the SW and that this effect remains active until the SW front curvature becomes small ($\theta$ = $1^{\circ}$, sec. \ref{sec:wavefront_forces}). We assume that we can take the vertical force felt by particles in our DSW wavefront to be $\mu\times sin(1^{\circ}) F_{m,3D}$, with the friction coefficient added as though it modifies the vertical force the same way that the $\mu$ modifies the normal force to find tangential force in our model (eq. \ref{eq:force_model}). It is not entirely clear how to include cohesion in this force balance and, though we've so far assumed that the cohesion in our simulation has a negligible impact on SID, this may not be the case when cohesion is higher than we've considered in this work (a particle-particle cohesive energy density of 1 kPa, which means bulk cohesion would be substantially less than on the Moon). For now we assume there is a differential cohesive force felt in the normal direction which modifies $F_{m,3D}$. The modification is a result of changing cohesive contact area ($A_c$ = $-\frac{\pi}{4}  (\delta^2 -4R\delta)$ for monodisperse particles) with the leading edge of the particle experiencing the initial overlap $\delta_0$ and the trailing edge experiencing the maximum overlap $\delta_m$. We will rearrange the Hertz spring force as we did in sec. \ref{sec:wavefront_forces} to write $\delta_m$ = $(F_{m,3D} / k)^{2/3}$ and $\delta_0$ is calculated from eqn. \ref{eq:overburden}. We designate these contributions as $F_{c,m}$ and $F_{c,0}$ corresponding to the cohesive force as a result of the maximum and initial overlaps.  Of course, this is not actually the case since the wavefront is not a single particle in width, but we use this approximation for this initial characterization. Finally, we can write eq. \ref{eq:overburden_Fm_balance}, which balances gravitational forces with SW wavefront forces. We can then solve for $z_{loft}$ over a range of varied parameters. We emphasize that this equation only predicts the last depth to which particles of a given size will experience a lifting force as a result of the SW, it cannot be integrated to get a an exact idea of the velocity profile of grains in the subsurface. Forces on grains where there is substantial curvature (i.e, closer to the surface than $z_{loft}$) are underpredicted by our small angle assumption. This could be resolved in future works by replacing the 1$^{\circ}$ assumption with an equation that predicts the angle. We also note that, since $\delta_0$ is also a function of depth (and is now included thanks to the cohesion terms), $z_{loft}$ must be solved numerically and we do so using \textit{MATLAB}. 

\begin{align} \label{eq:overburden_Fm_balance}
    \mu\times sin(1^{\circ})\times\left(F_{m,3D} - (F_{c,m} - F_{c,0})\right) = \nonumber \\ \left(\phi(\pi R^2)z_{loft} + \frac{4}{3}\pi R^3  \right)\times\rho_p g
\end{align}

\begin{figure}[h]
\begin{center}
\includegraphics[scale = 0.4]{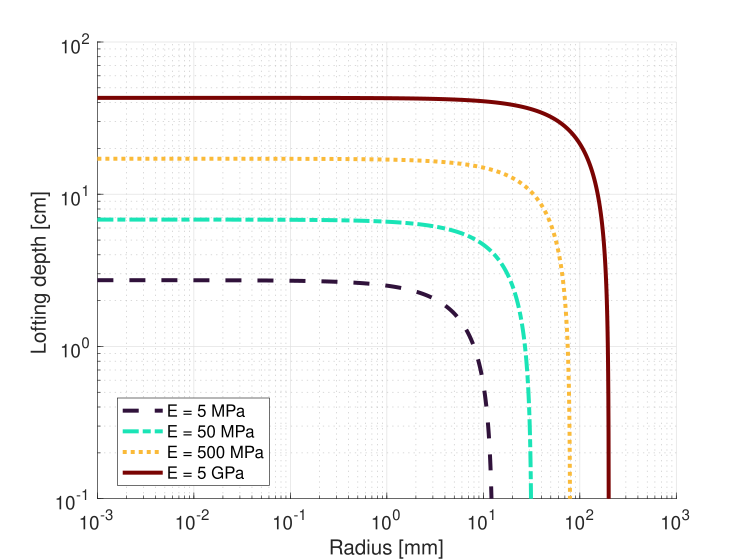}
\end{center}
\caption{\textbf{Loft depth vs maximum lofted particle size.} We show the lofting depth on the y axis which is the depth to which particles of a given size (radius, $R$, on the x axis) can attain an upwards velocity as the result of the SW wavefront.  }
\label{fig:loft_depth_scaling}
\end{figure}

Having assembled eq. \ref{eq:overburden_Fm_balance}, we can now inspect how loft depth changes with varied material parameters or gravitational loading. We solve for $z_{loft}$ for all particles larger than 1 micron (the smallest size particle in the lunar surface regolith profile from Fig. \ref{fig:particle_distributions}) for varied $E$, $\rho$, $k_c$, and $g$ using a fixed $v_0$ = 10 m/s (other parameters are set to the base material parameters, as usual). The solution is illustrated in Fig. \ref{fig:loft_depth_scaling} which shows how the $z_{loft}$ that is a result of SID changes with elastic modulus. While the plot is shown in log-log space to emphasize the order of magnitude increase in $z_{loft}$ for increasing $E$, note that the trend of $z_{loft}$ with $R$ is actually quite linear. The spring contribution from eq. \ref{eq:overburden_Fm_balance} dominates compared to the cohesive contribution. The x-crossing that is apparent in the plot corresponds to $R_c$, the critical particle radius above which no particles loft (the solution for $z_{loft}$ becomes negative and cannot be plotted in log-log space). Moving leftward along a given trend line from $R_c$ indicates the depth to which particles of a certain radius (assuming a monodisperse assembly) can acquire vertical velocity. As the trends approach the y-axis (from the right) they are increasing, though slowly compared to the increases that come from varying $E$. We consider $z_{loft}$ for particles of a given radius to be the loft depth ($z_{LD}$), that is, the deepest that SID can excavate the particles in each of our channels.

Since plots for the other parameters are similar (varying a parameter shifts the trends, but the shape is the same) we summarize the results of this same procedure to find $R_c$ and $z_{LD}$ across the parameter space in table \ref{tab:loft_depth_prediction_table}. We see that $R_c$ and $z_{LD}$ move together and increase for harder particles as expected (larger $E$ and $\mu$, smaller $R$). As the effective weight of the particles is reduced (i.e, decreased $\rho$ or $g$) the lofting parameters also increase and we should expect to see that gravity is the dominant factor. A one order of magnitude reduction in gravity (from the lunar 1.625 m/s$^2$ to 0.1 m/s$^2$) produces the same order of magnitude increase in $z_{LD}$ as a three order of magnitude increase in $E$ (from 5 MPa to 5 GPa). We expect to see very little effect on SID lofting from cohesion since the spring force is dominant in comparison. There is no change in $R_c$ even for the largest $k_c$ considered, though large $k_c$ does begin to decrease $z_{LD}$ for smaller particles. Cohesion plays a larger role in small particle dynamics (\cite{hartzell2011role}) so it makes sense that smaller particles see a large effect with increasing $k_c$, albeit still quite small in comparison to contributions from the other parameters. Note that our method of applying friction means that SID cannot occur for frictionless particles. In the next section we compare these predictions to our simulation results.

\begin{table}[h!]
\begin{minipage}{205pt} 
\caption{Loft depth ($z_{LD}$) and critical radius ($R_c$) for various material parameters and gravity environments. We solved eq. \ref{eq:overburden_Fm_balance} for $R$ in the case where $z_{loft}$ = 0 (the critical radius, $R_c$, which is the size of the largest particles that could be lofted by SID) and for $R$ = 1.25 mm (the size of the particles we used in our simulations for the cases in the table) to find lofting depth, $z_{LD}$, the depth to which we expect particles to obtain a vertical velocity (and therefore show bulk density changes). The parameters (besides the listed varied parameter) are set to the base properties, with $v_0$ = 10 m/s and $\phi_0$ = 62\%.} \label{tab:loft_depth_prediction_table}
\begin{tabular}{@{}cccc@{}}
\toprule
\begin{tabular}{@{}c@{}} Varied \\ Parameter \end{tabular} &
\begin{tabular}{@{}c@{}} Value \\ (units) \end{tabular} &
\begin{tabular}{@{}c@{}} $z_{LD}$ \\ (cm) \end{tabular} &
\begin{tabular}{@{}c@{}} $R_c$ \\ (mm) \end{tabular}  \\

\midrule

 $E$ & 5 (MPa) & 2.42  & 12.6  \\
 $E$ & 50 (MPa) & 6.53  & 31.7  \\
 $E$ & 500 (MPa) & 16.83  & 79.5  \\
 $E$ & 5 (GPa)  & 42.69  & 199.8  \\
  $E$ & 50 (GPa)  & 107.64  & 501.8  \\
 \hline
 $\rho$ & 0.8 (g/cc) & 3.97  & 19.9  \\
 $\rho$ & 3 (g/cc) & 2.23  & 11.7  \\
 $\rho$ & 10 (g/cc) & 1.28  & 7.2  \\
 $\rho$ & 25 (g/cc) & 0.81  & 5.0  \\
 \hline
 $g$ & 0.1 (m/s$^2$) & 43.74  & 204.8  \\
 $g$ & 1.625 (m/s$^2$) & 2.69  & 12.6  \\
 $g$ & 3.721 (m/s$^2$) & 1.18  & 5.5  \\
 $g$ & 9.8 (m/s$^2$) & 0.45  & 2.1  \\ 
 \hline
 $k_c$ & 0 (kPa) & 2.44  & 12.6  \\
 $k_c$ & 0.1 (kPa) & 2.44  & 12.6  \\
 $k_c$ & 1 (kPa) & 2.42 & 12.6  \\
 $k_c$ & 10 (kPa) & 2.25  & 12.6  \\
 \hline 
 $\mu$ & 0  & n/a  & n/a  \\
 $\mu$ & 0.4  & 0.81  & 5.0  \\
 $\mu$ & 1  & 2.42  & 12.6 \\
 $\mu$ & 1.8  & 4.58  & 22.7  \\

\bottomrule
\end{tabular}
\end{minipage}
\end{table}

\subsection{Dilation} \label{sec:dilation_results}

\begin{figure*}[h!]
\begin{center}
\includegraphics[scale = 0.21]{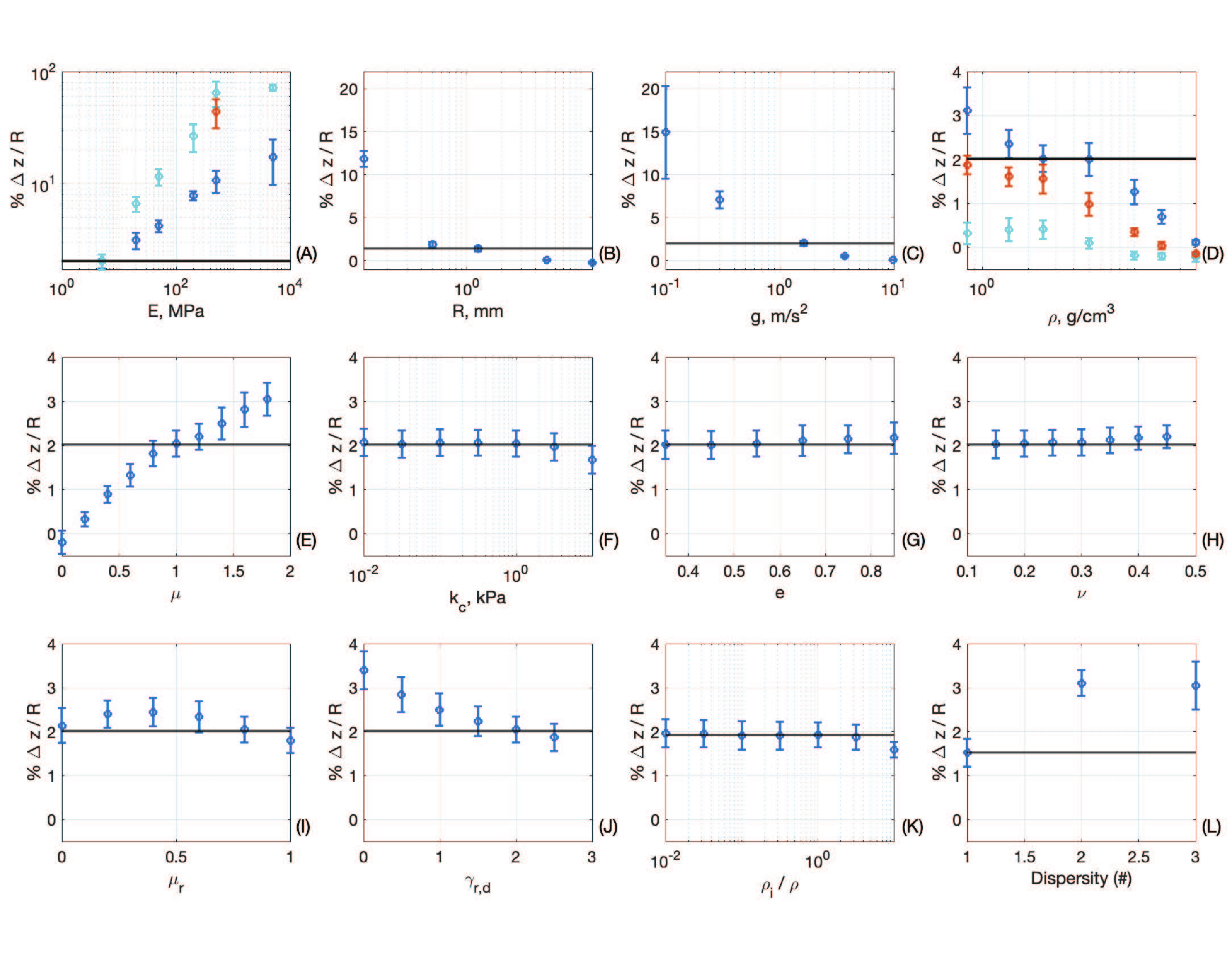} 
\end{center}
\caption{\textbf{Lofted height vs varied material parameters.} We show normalized percent change in bed height vs a sweep of various material and simulation parameters. The plots correspond to varied \textbf{A)} elastic modulus, \textbf{B)} radius, \textbf{C)} gravity, \textbf{D)} density, \textbf{E)} sliding friction , \textbf{F)} cohesion, \textbf{G)} restitution, \textbf{H)} Poisson's ratio, \textbf{I)} rolling friction \textbf{J)} rolling viscous damping \textbf{K)} momentum ratio and \textbf{L)} dispersity. In each case, the solid black line corresponds to percent bed height change of the reference case (generally a channel filled to 20 cm with particles of $R$ = 1.25 mm, $E$ = 5 MPa, and $\rho$ = 2.5 g/cc. \textbf{A)} Shows the case for constant $v_0$ = 10 m/s in blue and the case where $v_0$ is a constant portion of $c_m$ in cyan. The single red point corresponds to the lunar-like particles ($\rho$ = 3 g/cc, $R$ = 125 $\mu$m) with $v_0$ = 10 m/s. \textbf{D)} Shows bed height change vs particle density for various levels of $\phi_0$, with the compact, medium, and loose channels corresponding to blue, red, and cyan markers (respectively). \textbf{F)} The data point for $k_c$ = 0 was left off given the log nature of the x-axis, but showed the same bed height change as in cases with cohesion ($\Delta z$ $\sim$ 2\%).}
\label{fig:dilation_vs_depth_materialparams}
\end{figure*}

\begin{figure*}[h!]
\begin{center}
\includegraphics[scale = 0.75]{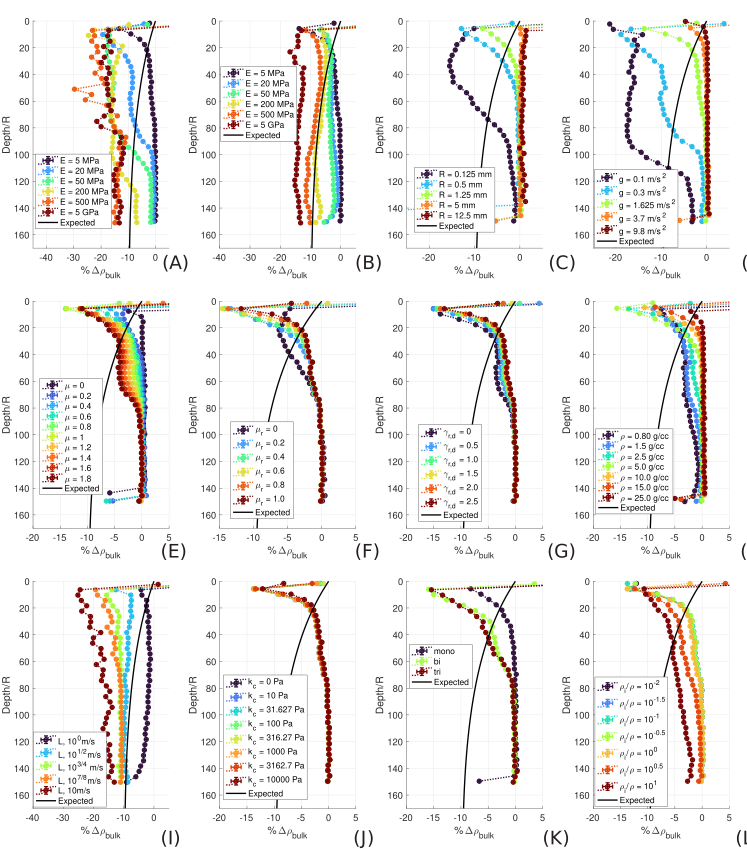}
\end{center}
\caption{\textbf{Dilation vs depth.} We show normalized depth ($z/R$) on the y axis vs percent bulk volume change on the x axis for cases of varying material parameters. Each plot has its own color legend corresponding to the varied parameter of interest. In all but the first case, only a single parameter is varied, with other parameters set to the base case from sec. \ref{sec:prefilled} and $v_0$ = 10 m/s. The first plot shows the case of varied $E$ and varied $v_0$, with $v_0$ held to the same portion ($\sim$20\%) of $c_m$. The plots correspond to varied \textbf{A)} elastic modulus and impact velocity, \textbf{B)} elastic modulus, \textbf{C)} $R$, \textbf{D)} gravity, \textbf{E)} sliding friction, \textbf{F)} rolling friction, \textbf{G)} rolling viscous damping, \textbf{H)} density, \textbf{I)} velocity (with lunar-like particles), \textbf{J)} cohesion energy density, \textbf{K)} dispersity and \textbf{L)} impactor momentum ratio. Note that the case for restitution was left off as it is essentially identical to the plot for $\nu$. In all cases the solid black line is the depth-normalized LCS expected bulk dilation from Fig. \ref{fig:bandfield_density_pdiff}.}
\label{fig:dilation_vs_depth_combo}
\end{figure*}

We now evaluate the SID demonstrated in our simulations. We examine the density change at depth for a range of different parameters and also measure the channel height change as a result of the DSW as was done in \cite{frizzell2023simulation}. As described in sec. \ref{sec:investigated_params}, we ran numerous tests while varying material parameters $E$, $R$, and $\rho$ as well as several important simulation parameters. To compare results across the parameter space, we evaluate the SID induced using a fixed $v_0$ of 10 m/s (except for one case, where we increased $v_0$ to be a constant portion of $c_m$ as $E$ increased). It is easiest to see how SID behaves by examining both the channel height change (Fig. \ref{fig:dilation_vs_depth_materialparams}) affected by the DSW wave as well as how bulk density changes with depth (Fig. \ref{fig:dilation_vs_depth_combo}).  We consider both plots together as we step through the different parameters evaluated. 

First, Fig. \ref{fig:dilation_vs_depth_materialparams} shows channel height change ($\Delta z$), with each subplot showing the percent change in channel height (normalized by $R$) from an initial 160$R$ height vs different material and simulation parameters. The single black line drawn across each subplot indicates the reference value which is the percent height change using the base parameters from \cite{frizzell2023simulation}. The percent height change axis is scaled similarly for most plots (in the range of -1 to 4 \%), but note the larger ranges shown for the first three plots for elastic modulus, particle size, and gravity which have the largest influence on SID (as we predicted in sec. \ref{sec:scaling_predictions}). Second, Fig. \ref{fig:dilation_vs_depth_combo} shows bulk density change ($\%\Delta\rho$) on the x axis for a given depth on the y axis, where the depth is normalized by particle radius. We also include the expected density change from \cite{bandfield2014lunarCS} as we initially described in Fig. \ref{fig:bandfield_density_pdiff} for reference. There seem to be two categories of SID: 1. a gentle dilation where $\%\Delta\rho$ decreases by `peeling away' near the surface, and 2. a more energetic rearrangement of particles where the entire channel lifts up and sets back down in a looser state following the DSW. For example, SID in cases where material properties are close to the base material parameters (softer particles, like in Fig. \ref{fig:dilation_vs_depth_combo}G) show gentle dilation while the hardest particles in the largest E cases (like those in  Fig. \ref{fig:dilation_vs_depth_combo}A) demonstrate an energetic rearrangement of the channel, with the grains falling back down to approximately the same (heavily dilated, $\%\Delta\rho\sim-15\%$) post SW packing fraction. Energetic rearrangement occurs when the impulse of a contact with the fixed $v_0$ traveling through the floor particles (discussed in sec. \ref{sec:wavegen}) delivers sufficient energy to lift entire depth of the channel at the contact point. The DSW wavefront effectively blasts the floor from below across the length of the channel. In the case of energetic rearrangement, we should find that the predicted $z_{LD}$ is larger than the channel height. We again emphasize that the depth range of our channels represents a small portion of the actual LCS depth. On the other hand, when dilation is gentler, forces on the particles give a gradual upward lift that increases density while leaving the initial arrangement of grains mostly intact. Gentle dilation is a result of floor impacts inducing a DSW which sees the force from particle collisions in the wavefront exceeding overburden forces only at some intermediate depth between the surface and the channel floor. For these cases in which our channel height encompasses the predicted $z_{LD}$, we can collect a simulated lofting depth by evaluating the `peel away' depth in Fig. \ref{fig:dilation_vs_depth_combo}. We will compare and contrast expected and predicted $z_{LD}$ shortly, but first we discuss qualitatively how SID is affected by changes to our investigated parameters.

\begin{figure*}[h]
\begin{center}
\includegraphics[scale = .63]{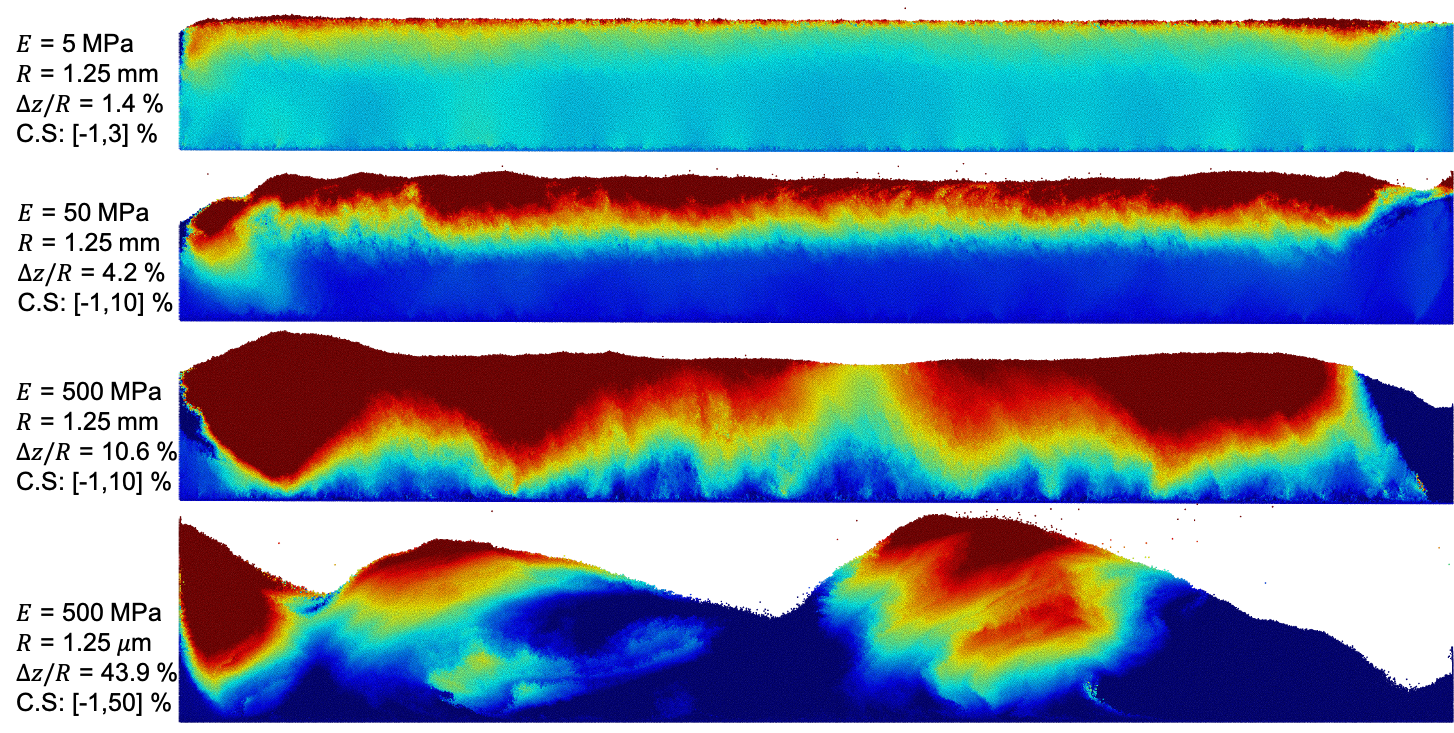}
\end{center}
\caption{\textbf{Dilated channels.} Particles are colored by the percent height change with respect to their initial (pre-SW) state. We use \textit{OVITO} to create the image and normalize particle vertical displacement by $R$, then find the percent difference compared to the initial 160$R$ channel height. We show the color scale (C.S) for each frame in a box immediately to the left of the post-SW channel, where darkest blue is represented by the lower value and dark red by the higher value. Each box also includes details on the particle $E$ and $R$ as will as the percent height change corresponding to Fig. \ref{fig:dilation_vs_depth_materialparams}. Frames 1-3 show channels with the base parameters and increasing $E$ while frame 4 shows the case corresponding to the lunar-like particles. All channels experienced a piston impact at $v_0$ = 10 m/s. Dilation is `gentler' for the softer particle channels which corresponds to uniform (and smaller in magnitude) dilation that is present in shallow surface depths. As the particles become harder the SID is `stronger' and most of the particles in the channel are thrown up and set back down in looser heaps.} 
\label{fig:dilatedbeds}
\end{figure*}

Considering the elastic modulus, Fig. \ref{fig:dilation_vs_depth_materialparams}A for varied $E$ shows both the case for $v_0$ held at a constant portion of $c_m$ as $E$ is varied (cyan) and the case where $E$ is increased while $v_0$ is constant at 10 m/s (blue). When $v_0$ is a constant portion of $c_m$, the two highest $E$ used (500 MPa, 5Gpa) produce approximately the same amount of dilation (nearly 100\%). This is a result of energetic rearrangement. Table \ref{tab:loft_depth_prediction_table} shows that the predicted lofting depth (42 or 107 cm corresponding to the two largest $E$) exceeds the channel depth we consider (20 cm for this channel). With the lofting depth exceeded, all particles in the channel are lofted as a result of the DSW, and they undergo a similar grain rearrangement, falling to the floor as though they were poured at approximately the same fast rate (`energetic rearrangement'). In other words, there is only so much material in the channel, it is all lofted in any case where $z_{loft}$ exceeds channel height, and the grains are rearranged to a similarly loose state. Smaller $E$ demonstrate the gentler SID response. Fig. \ref{fig:dilation_vs_depth_combo}B shows the dilation vs depth for when $E$ is increase but $v_0$ is held constant. While $v_0$ = 10 m/s is an impact velocity below $c_0$ in a channel filled of particles with $E$ = 5 GPa (see Table \ref{tab:validity}), we still see substantial dilation to the same levels induced by $v_0$ that increase as $c_m$ increases with larger $E$. This type of dilation (when $v_0$ driving the DSW through the floor is much less than $c_0$) causes `outbursts', where no DSW wavefront is front is formed, but surface dilation still occurs as a result of the impact from below. We will describe outbursts further in sec. \ref{sec:floor_vortex}. We point to Fig. \ref{fig:dilatedbeds} to see how SID increases with increasing $E$. The image transitions from gentle to energetic SID as $E$ increases (upper three frames) while the last frame shows the result of one of the most `energetic' cases considered, where increased $E$ coupled with decreased $R$ rearranges the entire channel.

Turning to our other parameters of interest, increases in particle size decrease SID while smaller particles have an easier time lofting as predicted in sec. \ref{sec:scaling_predictions}. In fact, for the largest particles ($R$ = 12.5 mm), SID is nonexistent as can be seen in both Fig. \ref{fig:dilation_vs_depth_materialparams}B and Fig. \ref{fig:dilation_vs_depth_combo}C. This agrees with sec. \ref{sec:scaling_predictions}: we predicted a critical radius of $R_c$ = 12.6 mm (see Table \ref{tab:loft_depth_prediction_table}) for particles with $E$ = 5 MPa (as is the case in Fig. \ref{fig:dilation_vs_depth_materialparams}B). We would expect almost no lofting, with any lofting that did occur taking place only at the surface. Note that the channel with $R$ = 12.5 mm particles had a slightly lower $\phi_0$ (it originally settled to a looser $\phi_0$, see sec. \ref{sec:variedR_fill}) which would reduce SID some (less than 1\% change between different $\phi_0$ cases at $\rho$ = 2.5 g/cc in Fig. \ref{fig:dilation_vs_depth_materialparams}D). However, particle size is dominant in setting the magnitude of SID (as compared to $\phi_0$) as evidenced by the 15\% $\Delta$z/R for $R$ = 125 $\mu$m in Fig. \ref{fig:dilation_vs_depth_materialparams}B.

Our predictions regarding gravity also agree with simulations. We predicted that a single order of magnitude decrease in gravity would show the same magnitude of SID as a three order of magnitude increase in $E$ and this is very nearly the case (compare percent height change in Figs. \ref{fig:dilation_vs_depth_materialparams}C and \ref{fig:dilation_vs_depth_materialparams}A along with frames Figs. \ref{fig:dilation_vs_depth_combo}B and \ref{fig:dilation_vs_depth_combo}D). We confirm that particle hardness (as seen by the Hertz spring constant k, which depends on E and R) and gravity are the main factors in determining if and when SID occurs.

Friction, density, and dispersity had a noticeable, though smaller, influence on SID. As predicted in Table \ref{tab:loft_depth_prediction_table} and confirmed in Figs. \ref{fig:dilation_vs_depth_materialparams}D and \ref{fig:dilation_vs_depth_combo}H, there is a small increase in SID with decreasing density and the heaviest particles are unable to loft. This makes sense; $F_m$ depends on $R^2$ but the mass depends on $R^3$, so increasing particle size leads to a faster increase in the overburden force than the wavefront force. Increasing friction could lead to a doubling in channel height change in the cases we considered (see Fig. \ref{fig:dilation_vs_depth_materialparams}E and Fig. \ref{fig:dilation_vs_depth_combo}E) and can be seen to slightly increase $z_{LD}$. Notably, the case with $\mu$ = 0 showed compaction, not dilation. Even in a channel with a large initial packing fraction ($\phi_0 \sim 62\%$), frictionless spheres cannot exhibit SID. Recall that, when friction is nonzero, the channel can also dilate if the initial packing fraction is low enough ($<$ 55\% for the monodisperse assembly in \cite{frizzell2023simulation}), though we did not investigate the crossover threshold in this work. Despite the bi- and tri- disperse channels filling to looser packing fractions than the monodisperse case, they saw nearly a doubling in channel height change (Fig. \ref{fig:dilation_vs_depth_materialparams}L) and the dilation vs depth profile for each are quite similar (Fig. \ref{fig:dilation_vs_depth_combo}K).

The remainder of the cases investigated either had a small (rolling viscous damping) or negligible affect on the expression of SID. Increased $\gamma_{r,d}$ decreased the channel height change slightly (Fig. \ref{fig:dilation_vs_depth_materialparams}J), but had little affect on the dilation vs depth profile in \ref{fig:dilation_vs_depth_combo}G. Rolling friction very slightly increased the extent of SID (see Fig. \ref{fig:dilation_vs_depth_materialparams}I around $\mu_r$ = 0.4), but mostly is responsible for giving our simulated bulk volume change vs depth its characteristic shape (Fig. \ref{fig:dilation_vs_depth_combo}F). Note how for the case of $\mu_r = 0$ the percent change in bulk density curve is missing the characteristic near surface peak in $\%\Delta\rho$ that `gentle SID' plots show.  The last case we can compare to the sec. \ref{sec:scaling_predictions} prediction is cohesion, for which we predicted there should be very little influence on SID. In agreement, Fig. \ref{fig:dilation_vs_depth_materialparams}F shows that there was only a small decrease in the amount of channel height change experienced when particles experienced the largest $k_c$ and the same is seen in Fig. \ref{fig:dilation_vs_depth_combo}J. Note that $\nu$ and $e$ had essentially no effect on SID (check out either Figs. \ref{fig:dilation_vs_depth_materialparams}G or \ref{fig:dilation_vs_depth_materialparams}H). It makes sense that $\nu$ has very little impact on SID since it only enters into eqn. \ref{eq:overburden_Fm_balance} by way of the equivalent properties (elastic modulus and radius) and was varied over a small range compared to the other parameters. It is interesting that $e$ has no effect on SID because reducing $e$ generally leads to greater energy dissipation \cite{manciu2001impulse}. SID's lack of dependence on $e$ suggests that the magnitude of the impact driven from below is the dominant factor in determining surface lofting compared to the plasticity of the grains within the vacuum-exposed channel.

Increasing impactor momentum by way of increasing the density in the piston particles had very little influence on SID, see Fig. \ref{fig:dilation_vs_depth_materialparams}K. This makes sense; we only altered the mass of piston particles (and thus the PSW), not those in the floor (in the DSW). Note that while the amount of total dilation seen by way of channel height change did not change when impactor momentum was varied in this way, the dilation vs depth in Fig. \ref{fig:dilation_vs_depth_combo}L shows a slightly different bulk density change vs depth profile. However, in calculating the total dilation, the increased dilation at deeper levels was canceled out by reduced dilation at the surface.  When momentum was increased by increasing $v_0$ SID increases (Fig. \ref{fig:dilation_vs_depth_phi}) since the floor particles sustaining $v_0$ (and therefore the DSW, and ultimately surface dilation) do experience this increase. We note that increasing impactor velocity slightly increased the length of the PSW (BL) region while increasing momentum has a larger influence (see Figs. \ref{fig:dm_and_d0_blastloading}, B and C in sec. \ref{sec:blast_loading_results}). Changing momentum by way of varying mass has a greater effect on the distance over which energy can be transmitted. 

Finally, we examined a single case of how SID changed when changing both elastic modulus and particle size. We show the results of a velocity sweep in a channel filled with lunar-like particles in Fig. \ref{fig:dilation_vs_depth_combo}I. Dilation for this channel (particles of $E$ = 500 MPa, which is still one to two orders of magnitude below lunar particles and $R$ = 125 $\mu$m, which is on the same order as surface regolith) easily exceeds the level expected, with impacts as low as $v_0$ $\sim$3 m/s inducing $\%\Delta\rho$ greater than 10\% over nearly the entire channel depth. Compared to most of the other cases inspected, the particles in the lunar-like channel (shown in the bottom frame of Fig. \ref{fig:dilatedbeds}) end up with highly uneven surface. The heaping behavior of this channel demonstrates that smaller particles more easily achieve `energetic SID'. However, characterizing the dynamics of this behavior requires analysis beyond the single case inspected here. To deal with the uneven surface we restricted the height change measurement shown in Fig. \ref{fig:dilation_vs_depth_materialparams}I to the `second heap' (average height change between 800$R$ and 1000$R$ in this 1600$R$ channel) which, at around 40\% $\Delta z / R$ (0.87 cm height change in a 2 cm channel!), exceeds that of a channel containing particles with the same $E$ but larger $R$. We point out that the substantial dilation in these cases occurred at low piston speed, when $v_0$ is less than $c_0$. As we saw previously for other channels with larger $E$, the SID induced with the lunar-like particle channel was was the result of outbursts as opposed to a DSW wavefront and will be discussed further in sec. \ref{sec:floor_vortex}.

\subsubsection{Comparison to prediction} \label{sec:dilation_comparison}
We now evaluate how our predictions compare to simulated results for a subset of parameters in Table \ref{tab:loft_depth_prediction_table}. We manually collect the lofting depth from our simulations ($z_{SD}$, `simulated depth') for the cases in Fig. \ref{fig:dilation_vs_depth_combo} that show gentle SID and have $z_{SD}$ that is identifiable as the depth which $\%\Delta\rho$ exceeds 1\% dilated. These gentle SID cases should  correspond to predictions from Table \ref{tab:loft_depth_prediction_table}  that have $z_{LD}$ less than channel height. We present the results in Table \ref{tab:loft_depth_loft_height_comp} and compare $z_{SD}$ to $z_{LD}$ and $\Delta z$ representative measures of SID. 

\begin{table}[h!]
\begin{minipage}{220pt} \label{tab:z_loft}
\caption{Prediction vs comparison table. We find $z_{SD}$ for cases that exhibit gentle SID. We manually identified the depth at which $\%\Delta\rho$ exceeds 1\% by examining the different frames of Fig. \ref{fig:dilation_vs_depth_combo}, multiplying the depth identified as $z_{SD}$ by the particle $R$ for the corresponding case to recover dimensional units. A value of 0 indicates no dilation or compaction and a value of n/a means that there was no identifiable $z_{SD}$ (e.g., the lofting depth greater than the channel depth). We then compare $z_{SD}$, $z_{LD}$, and bed height change.  } \label{tab:loft_depth_loft_height_comp}
\begin{tabular}{@{}cccccc@{}}
\toprule
\begin{tabular}{@{}c@{}} Varied \\ Param. \end{tabular} &
\begin{tabular}{@{}c@{}} Value \\ (units) \end{tabular} &
\begin{tabular}{@{}c@{}} $v_p$ \\ (m/s) \end{tabular} &
\begin{tabular}{@{}c@{}} $\Delta z$ \\ (cm) \end{tabular} &
\begin{tabular}{@{}c@{}} $z_{LD}$ \\ (cm) \end{tabular} &
\begin{tabular}{@{}c@{}} $z_{SD}$ \\ (cm) \end{tabular}  \\

\midrule

 $E$ & 5 (MPa) & 10.0  & 0.410  & 2.42  & 9.1   \\
 $E,v_0$ & 20 (MPa) & 20.0  & 1.319  & 10.55   & 12.8   \\
 $E,v_0$ & 50 (MPa) & 31.623  & 2.313  &  26.82  & 15.3    \\
 $E$ & 20 (MPa) & 10.0  & 0.624  & 4.44   & n/a   \\
 $E$ & 50 (MPa) & 10.0  & 0.838  & 6.53  & n/a    \\
 $E$ & 200 (MPa) & 10.0  & 1.564   &  11.58  & n/a    \\
 $E$ & 500 (MPa) & 10.0  & 2.128  & 16.83  & n/a    \\
 \hline
 $\rho$ & 0.8 (g/cc) & 10.0 & 0.624  & 3.97  & n/a    \\
 $\rho$ & 2.5 (g/cc) & 10.0 & 0.404  & 2.42  & 9.1    \\
 $\rho$ & 10.0 (g/cc) & 10.0 & 0.252 & 1.28  & 5.6    \\
 $\rho$ & 25 (g/cc) & 10.0  & 0.023  & 0.81  & 1.95    \\
 \hline
 $g$ & 1.625 (m/s$^2$) & 10.0  & 0.410  & 2.69  & 9.1    \\
 $g$ & 3.721 (m/s$^2$) & 10.0  & 0.109  & 1.18  & 5.7    \\
 $g$ & 9.8 (m/s$^2$) & 10.0  & 0.023  & 0.45  &1.5    \\ 
 \hline
 $k_c$ & 0 (kPa) & 10.0  & 0.411  & 2.44  & 9.1    \\
 $k_c$ & 1 (kPa) & 10.0  & 0.410  & 2.44  & 9.1    \\
 $k_c$ & 10 (kPa) & 10.0  & 0.335  & 2.25   & 9.1   \\
 \hline 
 $\mu$ & 0.2  & 10.0  & 0.065  & 0.27  & 5.7     \\
 $\mu$ & 0.6  & 10.0  & 0.264  & 1.35  & 8.7    \\
 $\mu$ & 1  & 10.0  & 0.410   & 2.42  & 9.1    \\
 $\mu$ & 1.4  & 10.0  & 0.499  & 3.50  & 13.2    \\
 $\mu$ & 1.8  & 10.0  & 0.610  & 4.58  & 15.7    \\
 \hline 
  $R$ & 125 ($\mu$m)  & 10.0  & 0.237  & 2.66  & n/a    \\
 $R$ & 0.5 (mm)  & 10.0  & 0.150  & 2.58  & 14.6     \\
 $R$ & 1.25 (mm)  & 10.0  & 0.277  & 2.42  & 9.1    \\
 $R$ & 5 (mm)  & 10.0  & 0.057  & 1.62  & 0.7    \\
 $R$ & 12.5 (mm)  & 10.0  & -0.482  & 0.002  & 0.0    \\
 
\bottomrule
\end{tabular}
\end{minipage}
\end{table}

The expected trends hold true for all three quantities across all parameters; SID increases with increasing $E$, increasing $v_0$, increasing $\mu$, decreasing $R$, decreasing $\rho$, and decreasing $g$. Cohesion has a negligible influence on SID except for all but the largest $k_c$ we considered, which still has only a small influence on SID compared to the other parameters. The total height change of the channel is typically an order of magnitude less than the lofting depth. Comparing $z_{LD}$ to $z_{SD}$, our predicted lofting depth is, in most cases, several factors smaller (3-4$\times$) than our simulated lofting depth. The discrepancy indicates that our method of solving eqn. \ref{eq:overburden_Fm_balance} under-predicts SID, but the closeness of the results and the agreement across multiple orders of magnitude in parameter space given the numerous assumptions we made is a good indicator that we have correctly identified the underlying mechanism. Others investigating SID could improve the accuracy of prediction in the future by relaxing our assumptions. 

There are a few outlier cases to discuss. For the cases considering varied $E$, we would expect the loft depth in channels of increased $E$ and increased $v_0$ to quickly exceed the depth of the channel, and they do. However, we also see that when $v_0$ is fixed at 10 m/s and $E$ is increased, the simulated loft depth exceeded the channel height while the predicted loft depth only increased slightly (from 0.4 cm to 2.1 cm). These larger $E$ beds with smaller velocities were dilated by the outbursts we mentioned previously, as opposed to a DSW wavefront. Our prediction of the lofting depth is therefore only valid when it is a sustained uniform wavefront causing dilation. We will briefly discuss the behavior of outbursts in sec. \ref{sec:floor_vortex}, but future work should consider methods to predict when outbursts occur and to what extent they influence surface dilation. Nonetheless, our results have repeatedly shown that dilation increases as the particles in our channels become more lunar-like which is strong evidence that SID plays a significant role in the formation of LCS. While the method we have developed for predicting SID via loft depth includes numerous assumptions (that a fixed channel height is representative of any channel height, that there is a single scaling constant $C_1$ which is not influenced by material parameters itself, that the particles experience $\delta_0$ = 0, that the influence of cohesion is simply the difference between forces generated between contact areas set by $\delta_m$ and $\delta_0$) and does not take into account the outbursts that form at large $E$ but low $v_0$, it can still be used to determine the order of magnitude of the expected SID response, keeping in mind that it is a lower bound.

\subsection{Blast loading} \label{sec:blast_loading_results}
\begin{figure*}[h!]
\begin{center}
\includegraphics[scale = .58]{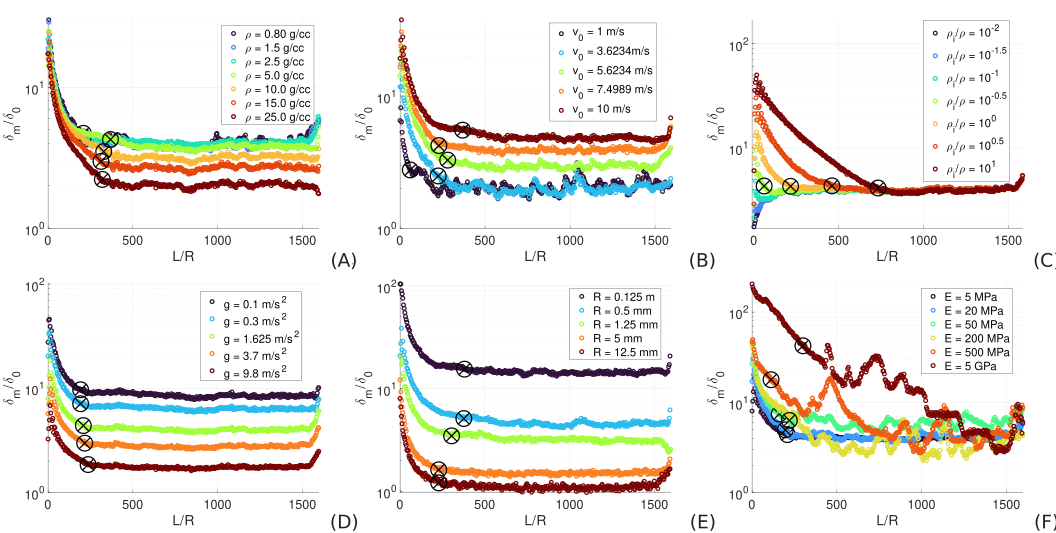}
\end{center}
\caption{\textbf{Blast loading via $\delta_m$.} We show the normalized maximum overlap in the SW wavefront ($\delta_m$/$\delta_0$) vs channel length normalized by particle size ($L$/$R$) for varied properties. The color legends refer to varied \textbf{A)} density, \textbf{B)} impact speed, \textbf{C)} momentum ratio, \textbf{D)} gravity, \textbf{E)} radius and \textbf{F)} E (for the case with constant $v_0$ = 10 m/s). When applicable, we indicate the end of the BL zone (end of the PSW, the dissipative region near the left wall) as the black crossed circle. We find the end of the BL zone using the procedure outlined in sec. \ref{sec:blast_loading_method} (when $\delta_m$/$\delta_0$ becomes close to the average steady value in the DSW). The material parameters have very little effect to the end of the BL zone, which is generally between 200 and 300 $R$. Changes to momentum have more of an effect on the lateral extent of the BL zone, with increased velocity (frame B) increasing the length of the region somewhat, while changes to momentum by tuning the density of the impactor particles (frame C) could either extend the region quite a bit (out to 800$R$) or result in no identifiable BL zone for low impactor density. Because we track $\delta_m$ as the maximum $\delta$ experienced by a particle (as opposed to just the $\delta$ experienced by the initial wavefront), plot 6 for varied $E$ is capturing effects from outbursts when $E$ is large. } 
\label{fig:dm_and_d0_blastloading}
\end{figure*}

One thing to be verified is that the blast loading region (region over which the PSW is active) in our simulations behaved as anticipated. We expect the DSW wavefront to overtake the PSW close to the piston initiation site (in terms of lateral distance) to allow for a sufficient length of channel over which to measure the DSW that is unimpeded by the end wall. We found good agreement with expectation and the BL zone varied little with particle properties. Increasing momentum of the impact (larger $v_0$ or larger $\rho_i$) had the largest influence on altering BL. Fig. \ref{fig:dm_and_d0_blastloading} visualizes the BL region vs the DSW region by tracking $\delta_m$ over the lateral extent of the channels. Channels in which only non-momentum affecting parameters were varied (frames A ($\rho$), D ($g$) and E ($R$)) show little variation in BL. There was a slight increase in BL with increasing density, and decreasing particle size, though mostly the BL region sizes were in the range of 100 to 200 particle diameters. Channels experiencing increased impact momentum (frame B ($v_0$) and C ($\rho_i/\rho$)) showed larger changes in BL (as is expected from blast loading tests in a 1D chain  \cite{hasan2018shock}), with the increased piston density case having the largest effect. The last image in frame F shows the case of varied $E$ but fixed $v_0$ = 10 m/s, which demonstrates further evidence of outbursts. Since we track $\delta_m$ as the maximum overlap experienced (not just in the wavefront), the cases with large $E$ are dominated by the $\delta$ within the outbursts and no longer resemble the usual PSW decay to driven SW pattern. The BL identified for the two cases of largest $E$ and fixed $v_0$ are left out of the averaging when computing the mean and deviation BL for each channel (summarized in Table \ref{tab:blastloadingtable}). In all cases (besides those affected by outbursts), the BL zone is shorter than our 1600$R$ channel length, confirming the validity of our channel size scaling choice (see sec. \ref{sec:variedR_fill} and \ref{sec:blast_loading_method}).

\begin{table}[h]
\begin{minipage}{205pt} 
\caption{Blast loading table. We give the mean ($\overline{BL}$) and deviation ($\sigma BL$) length of the BL zone for various cases. Each entry represents several tests conducted with the base particle properties and with the indicated test parameter varied over the range from sec. \ref{sec:investigated_params}. Units are given in non dimensional channel length ($L$/$R$). The velocity sweep case for $v_0$ comes from a channel with $R$ = 0.5 mm particles. The table is sorted by smallest deviation to largest.} \label{tab:blastloadingtable}
\begin{tabular}{@{}ccc@{}}
\toprule
\begin{tabular}{c} Test Parameter \end{tabular} &
\begin{tabular}{c} $\overline{BL}$ ($L$/$R$) \end{tabular} &
\begin{tabular}{c} $\sigma$BL ($L$/$R$) \end{tabular}  \\

\midrule

 $e$ & 208.0    & 0   \\
 $\nu$ & 208.0    & 0   \\
 $k_c$ & 208.0    & 0   \\
 $E,v_0$ & 197.3 & 8.3    \\
 $\gamma_{r,d}$ & 200.7    & 13.0  \\
 $g$ & 208.8    & 18.4   \\
 $\mu_r$ & 200.0  & 19.6  \\
 $\mu$ & 224.0    & 21.5  \\
 $E$ & 199.3    & 63.0   \\
 $\rho$ & 278.9   & 68.6   \\
 $R$ & 300.8  & 75.2  \\
 $v_0$ & 232   & 112.3  \\
 $\rho_i/\rho$ & 367.0    & 293.8   \\
 
\bottomrule
\end{tabular}
\end{minipage}
\end{table}

\subsection{Outbursts} \label{sec:floor_vortex}
\begin{figure}[h]
\begin{center}
\includegraphics[scale = .37]{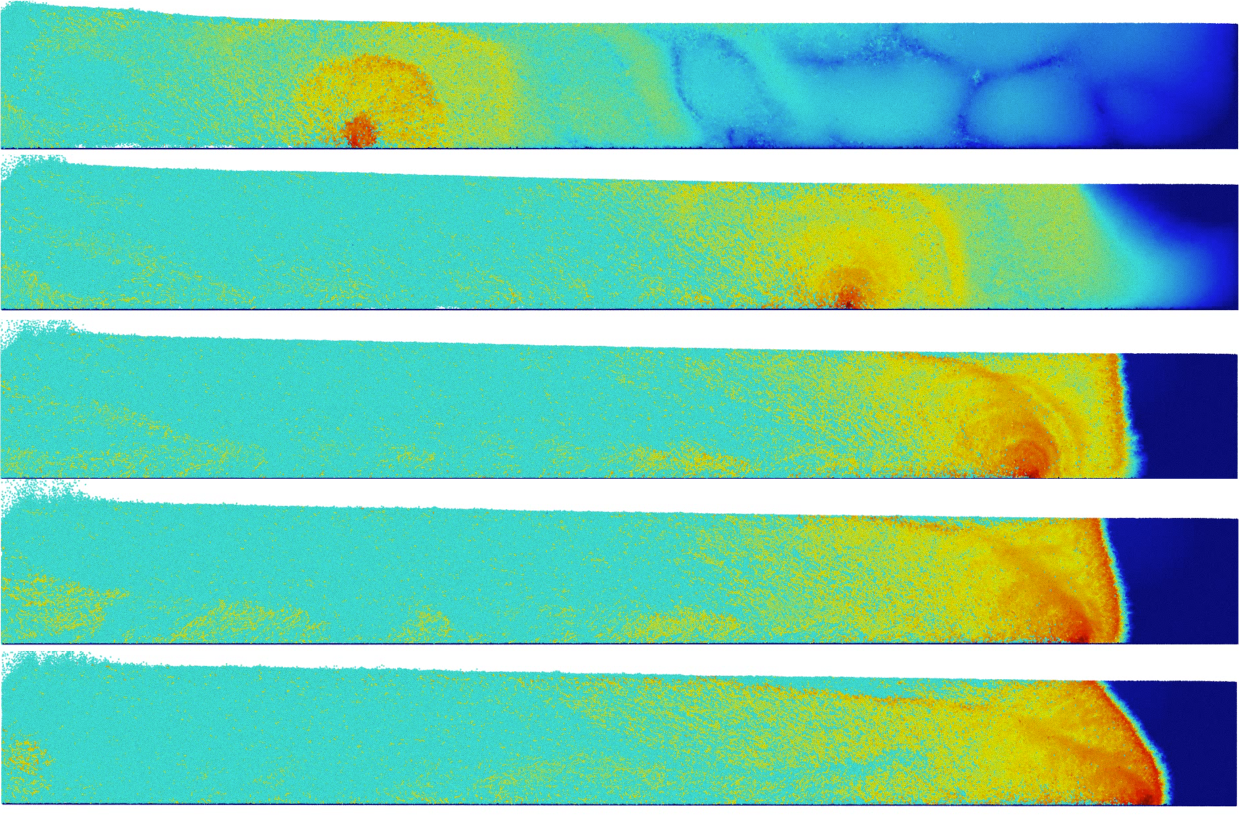}
\end{center}
\caption{\textbf{Outburst vs DSW.} Particles are colored by their force magnitude experienced relative to a reference force of 10$^{-6}$ N. The color legend is the same for all frames with the darkest blue at -2 (meaning the force a dark blue particle experiences is 2 orders of magnitude below the reference force) and the darkest red at 7. The images show the same channel (base parameters, but with $E$ set to 500 MPa). From Table \ref{tab:validity}, the sound speed for this channel is $c_0$ = 40.2 m/s. Each frame represents a different value of the initial piston velocity used, with the frames corresponding to (from top to bottom) a $v_0$ of \textbf{1)} 10 m/s, \textbf{2)}  31.6 m/s, \textbf{3)}  56.3 m/s, \textbf{4)} 74.9 m/s and \textbf{5)}  100 m/s (which is approximately 20\% of $c_m$). Frames are taken at different (arbitrary) times and and are selected so that the leading wavefronts (approximate PSW in frames 1 and 2 vs DSW in the others) are aligned. In frames 3-5, the energy delivered through any impact with the driven floor (including glancing oblique impacts) are large enough that a single well defined SW wavefront is created and maintained across the lateral extent of the channel. In frames 1 and 2 the conditions (geometrical arrangement of particles) that provide sufficiently large impact energies lead to occasional outbursts which, while they do not generate a coherent wavefront, are still able to dilate surface particles.} 
\label{fig:vortex}
\end{figure}

We have seen two distinct categories of granular wave behavior that occur when externally driving impacts through a channel using our perfect floor impact method (sec \ref{sec:wavegen}). When $v_0$ is much larger than the $c_0$, a single DSW wavefront propagates through the channel, but when $v_0$ is near or less than $c_0$ the result of the driven impacts are periodic outbursts of energy. We show a transition between the two categories in Fig. \ref{fig:vortex}, where the bottom images show a clearly driven SW wavefront while the top images show the outbursts as they pulse slowly behind the broken apart wavefront of the PSW. Outbursts are capable of exciting SID on the same order of magnitude as the DSW when particles are harder (through greater $E$) or smaller (lower $R$) as we saw in sec. \ref{sec:dilation_results}. Understanding differences in the dynamics of the two mechanisms requires further study since here we have primarily studied the influence of varied $v_0$ within the driving layer. The properties of the driving layer itself are largely neglected since those particles are `frozen' in our simulations. Since the floor layer is made up of a sheet of randomly packed particles, we believe the periodic nature of the outbursts to be correlated with the geometry of the packing. We suspect that outbursts occur when the geometry of particles is such that a force chain responsible for propagating the velocity through the floor makes a nearly normal (anti-gravity direction) impact with the channel particles. Then, when $E$ is larger or $R$ is low (or both), the energy delivered through this impulse is sufficiently large to propagate to and dilate grains at the surface. While both the DSW and outbursts induce SID, they rely on an external driver from below. In our simulations this is accomplished with the frozen-in floor velocity, which constantly adds energy through floor impacts to drive the SW (or outbursts). This raises questions about the possible external driver of the surface dilating wave which we leave for the Discussion.

\subsection{Discussion} \label{sec:discussion}

There were two main objectives in this paper; 1) develop a method to determine the maximum force $F_{m,3D}$ felt by a particle within a driven 3D SW wavefront and 2) determine how the magnitude of SID changes as that maximum force is scaled to granular mediums with different material properties. Remarkably, the results of our scaling analysis in sec. \ref{sec:wavefront_force_fitting} and Fig. \ref{fig:fm_vs_trends} show that $F_{m,3D}$ follows the same power laws that would be experienced by a particle within the SW of a 1D granular chain (eq. \ref{eq:Pal_FSW}), only at a reduced magnitude. We have seen that there is some variability in the exact value of the scaling parameter $C_1$ in this work ($C_1$ = 1-10$\times$10$^{-3}$), which may indicate that $C_1$ itself depends on material properties. Furthermore, $F_{m,3D}$ does not currently capture depth-dependent effects present in the SW wavefront, like the larger than expected forces near the surface (Fig. \ref{fig:fm_vs_depth_R}). Since our method does not capture these near surface effects, the value of $C_1$ we determined only applies to granular channels of 160$R$ in height and would change for taller (smaller $C_1$) or shorter (larger $C_1$) channels. Keeping these limitations in mind, we then proceeded to use $F_{m,3D}$ in a scaling analysis of SID, assuming that the influence of the accuracy of $C_1$ on our prediction is negligible in comparison to the power law scaling of the material parameters in $F_{m,3D}$.

To address our second goal, we developed a method (equating the wavefront force to the overburden force) that quantifies how SID changes with different material parameters and in different gravitational environments (eq. \ref{eq:overburden_Fm_balance}). Solving eq. \ref{eq:overburden_Fm_balance} for $z_{loft}$ = 0 gives the $R$ of the largest particles which could be lofted by SID ($R_c$). Solving eq.  \ref{eq:overburden_Fm_balance} for $z_{loft}$ given some $R$ gives the depth to which we'd expect those particles to show a dilation response ($z_{LD}$) as a result of the driven SW wavefront. An increase or decrease in $z_{LD}$ corresponds to increased or decreased SID. Fig. \ref{fig:dilation_vs_depth_combo} shows that $\%\Delta z$ increases as $z_{LD}$ increases, which also results in larger channel height changes ($\%\Delta z$), see Table \ref{tab:loft_depth_loft_height_comp}. In this work we did not attempt to quantify the velocity profile of grains between the surface and $z_{LD}$ and we are therefore unable to predict how their initial ballistic trajectories result in their rate of return to the surface (which controls $\%\Delta\rho$ and final channel height). Removing the small angle assumption and determining how to calculate the wavefront angle should enable this ability, which would be an interesting topic for future work. While our predicted loft depth was less than the simulated loft depth, they are still on the same order of magnitude and scale similarly in most cases. Our method of predicting SID can serve as a lower bound when predicting the extent of SID on low gravity and airless bodies. As expected, SID increases as particles become smaller and harder and SID decreases with increasing gravity. We are confident in the validity of our results given that particles within our channels that look closest to the lunar regolith undergo impact velocities within the SW that remain close to the elastic limit and do not experience overlaps that exceed 1\% of $R$. There is, however, some disagreement with prediction in a few cases (low $v_0$ impacts into channels with large $E$), where the lofting depth increased faster than expected or channels dilated when we did not expect them to. This discrepancy is the result outbursts which roll along the floor slower than a sound wave, but periodically pulse to generate a high force concentration that induces surface dilation.

While we do expect the mechanisms studied here to produce the magnitude of dilation needed to explain LCS when grains are lunar-like, the long range expression of SID in \cite{frizzell2023simulation} was artificial. Thanks to the fixed $v_0$ propagating through the floor in \cite{frizzell2023simulation} we could maintain DSW or outbursts in our simulation over any distance. The source driving floor impacts in our simulation is not real; it is the result of immovable particles (of the same size as the channel particles) that experience a $v_0$ which propagates without decay. However, we believe this situation does have a physical analog in the case of two distinct assemblies of grains in contact with one another, where the surface most layer experiences time variant energy injection from a buried layer. We consider an upper assembly (A1) of grains resting on top of and made up of grains that are several orders of magnitude smaller than a lower, buried, assembly of grains (A2). Recall that, in granular systems, energy decay is generally a function of particle number which is why smaller grains are more efficient impact absorbers (smaller grains occupy a smaller lateral extent than the same number of larger grains). Therefore, if our entire combined assembly experiences a uniform lateral impulse, energy would propagate over a much greater physical extent in the buried assembly (A2) made up of larger grains than in A1. If the particles in A2 are large enough compared to the particles in A1, then particles in A2 may experience sufficient velocities in its decaying SW to generate impacts on the A1 layer.  Particle velocities in A2 would decrease as its SW decays which suggests that, if this situation were to actually occur, an initially driven SW in A1 (as a result of higher impact velocities) would uniformly dilate the surface over some finite distance until lower particle velocities as a result of decay in A2 lead to outbursts in A1. This hypothetical scenario agrees well with the pattern of the LCS halo, where the surface is strongly and uniformly dilated (possibly due to a driven SW) around the source crater (up to $\sim$100 crater radii), transitioning to a patchy region of intermittent reduced thermal inertia (possibly due to outbursts) that extends beyond the well defined halo ($>$100 crater radii, see images in \cite{bandfield2014lunarCS} or \cite{hill2017well}). Moreover, our hypothetical scenario is based on the actual lunar near surface environment. LCS exist in the uniform micron sized grains of the regolith (a $\sim$10 m thick layer, \cite{venkatraman2023statistical}) which rests atop a layer of (probably) meter sized boulders known as the upper megaregolith (\cite{thompson1979blocky}, \cite{mckay1991lunar}, \cite{richardson2020modeling}). We encourage readers to consider the previous hypothetical scenario with A1 representing regolith and A2 the upper most portion of large scale ejecta in \cite{horz1991lunar}, Fig. 4.22. It is typically assumed that there is an abrupt transition between these two layers, as is the case in \cite{kovach1973velocity} when considering sound speed changes between the regolith and megaregolith. The scenario we considered here would support this sudden transition (as opposed to a gradient) between layers. This suggests that LCS are formed as a consequence of energy laterally propagated through a layer of large grains buried underneath a thick sheet of much smaller grains, and are a previously unconsidered process of surface modification as a result of impact induced seismicity on low-gravity airless-bodies.

While there is still more work to be done in understanding the precise impulse-induced lofting mechanism, the extent of dilation we observed in our simulations is strong evidence that such a mechanism is heavily involved in LCS generation. Dilation increases as particles become both smaller (lower $R$) and as they become harder (higher $E$). In the case we examined that is closest to lunar particles ($E$ = 500 MPa, $R$ = 125 $\mu$m, $\rho$ = 3 g/cc), we see the channel is completely rearranged following the passage of either the SW wavefront or the floor vortex. Our lunar-like channel exhibited a bed height change that is equivalent to $\sim$44\% dilation (Fig. \ref{fig:dilation_vs_depth_materialparams}, frame 1 for the percent height change and Fig \ref{fig:dilation_vs_depth_combo} to see how the lunar-expected dilation is exceeded) with $v_0$ of only 10 m/s. We expect to see dilation responses in our channels greater than is predicted for LCS since we consider only a fraction of the actual LCS channel height and shorter channels experience faster wave speeds for the same impact speed (\cite{frizzell2023simulation}, \cite{somfai2005elastic}). Our channels are 160$R$ in depth while an LCS channel exceeds 1000$R$ (at least in terms of the depth extent of an LCS over which dilation is observed, 40 cm). Moreover, the need for an external driver along a subsurface floor suggests that the entire regolith is the LCS channel, a channel height on the order of 10 m. Regardless of the actual lunar channel dimensions, the 2 cm bed height of our lunar-like particles is much less than the 40 cm LCS dilated region and so our results for low speed impacts that show dilation magnitudes exceeding expectation would be reduced as channel height increases. Future work should consider larger assemblies that more closely resemble the LCS scenario, both with taller channels, but also with lower packing fractions. The $\phi_0$ we considered in this work (around 58\%) are larger than what is found for lunar surface regolith ($\phi_0$ of 47 - 54\%, \cite{carrier1991physical}). Lower $\phi_0$ can be achieved by increasing the dispersity of the particle distribution as we saw in sec. \ref{sec:variedR_fill} (the same filling procedure for mono, bi-, and tri-disperse channels produced $\phi_0$ of 59, 54 and 52 \%, respectively). Although these modifications we describe would reduce the magnitude of dilation as compared to the cases modeled in this work, we believe they provide further support for SID as the LCS formation mechanism. We saw energetic SID in this paper for lunar-like particles, but it is the more gentle SID that would be involved in LCS creation given the dearth of signs of surface modification within the LCS halo.

\section{Conclusions}
In this work we have formulated a method for determining the maximum force a single particle feels as a result of a SW wavefront driven by constant impact with a layer of particles from below. We validated this method by demonstrating that the forces experienced in 3D follow the same scaling laws as in a 1D chain. Using our force prediction, we then constructed a method for analytically characterizing the expression of SID, which showed good agreement with numerical results (on the same order of magnitude). The primary results of our work are:

 \begin{itemize}
     \item the driven SW wavefront forces follow the same scaling power laws in 1D and 3D granular assemblies with a somewhat constant scaling between 1-10$\times$10$^{-3}$ (for channels filled with monodisperse particles to a depth of 160$R$)
     \item SID can be predicted using an equation to balance wavefront forces and gravitational overburden by assuming that the portion of the force from the wavefront felt in the vertical direction is reduced as a component of a small wavefront angle
     \item SID is primarily affected by the elastic modulus and particle size of grains as well as the gravitational environment:
     \begin{itemize}
     	\item SID increases the most with increasing $E$, decreasing $R$ and decreasing $g$
	\item SID increases some with decreasing $\rho$, increasing $\mu$ and increasing dispersity
	\item SID sees little influence from $k_c$, $e$, $\nu$, $\mu_r$ and $\gamma_{r,d}$ 
     \end{itemize}
     \item Impacts at $v_0$ $<$ $c_0$ generate periodic outbursts that cause similar magnitudes of surface dilation as those seen from a single, coherent, driven SW wavefront
 \end{itemize}

We expect granular assemblies that have material properties similar to that of lunar surface regolith to produce the same level of bulk dilation as is expected in the production of Lunar Cold Spots. We suspect that, if involved in LCS formation, the driven SW and outbursts we describe here would be maintained in micron scale surface regolith by the decay energy of an impact propagating through an underlying layer of nearly continuous meter scale boulders within the upper megaregolith.

\backmatter

\bmhead{Supplementary information}
We have generated numerous videos of our simulations that help to visualize SID processes. Videos from velocity sweeps (varied $v_p$) in channels filled with particles of varied $E$ and $R$ which correspond to cases A, B and C from Fig. \ref{fig:dilation_vs_depth_combo} as well as videos from simulations conducted in the channel filled with lunar-like particles (case I, Fig. \ref{fig:dilation_vs_depth_combo}) are provided at CITEME. File names indicate videos taken to show either the initial shock (\textit{shock}), the driven solitary wave (\textit{dsw}) or dilation (\textit{dilation}). Outbursts can be seen in any \textit{dilation} video where $v_p$ $<$ $c_0$. We have selected a few of the most salient cases to provide as supplementary material directly attached to the online version of this article. We demonstrate the effects of piston impact into channels corresponding to cases OL1: particles of $R$ = 125 $\mu$m, $E$ = 5 MPa, $\rho$ =  2.5 g/cc at $v_p$ = 10 m/s (example of dilation due to gentle SID), OL2: particles of $R$ = 1.25 mm, $E$ = 500 MPa, $\rho$ =  2.5 g/cc at $v_p$ = 100 m/s (example of dilation due to energetic SID), OL3: $R$ = 125 $\mu$m, $E$ = 500 MPa, $\rho$ =  3.0 g/cc at $v_p$ = 10 m/s (example of outburst propagation in the lunar-like channel), and OL4: $R$ = 125 $\mu$m, $E$ = 500 MPa, $\rho$ =  3.0 g/cc at $v_p$ = 1 m/s (example of dilation due to outbursts in the lunar-like channel).

\bmhead{Acknowledgements}
We thank Y. Zhang for helpful granular dynamics discussions and pointing us to the Voronoi tessellation tool. Thanks as well to the University of Maryland High Performance Computing and Division of Information Technology for their support.

 \section*{Declarations}
 
\begin{itemize}
\item Funding: None.
\item Competing interests: The authors have no competing interests to declare that are relevant to the content of this article.
\item Ethics approval: Not applicable.
\item Consent to participate: Not applicable.
\item Consent for publication: Not applicable.
\item Availability of data and materials: Given the size of our datasets we provide a subset of data used in this paper. The restart files and scripts needed to launch shock simulations along with said restart files are provided in UPDATE\cite{frizzell2023data}. We also provide all the output from a single shock run. Any reader could use the provided files with open source software (see Code availability below) to easily recreate our work and validate against the provided single shock run.
\item Code availability: We use open source code for simulation and visualization. The files used to implement our model and run the simulations are provided in our public repository UPDATE(\cite{frizzell2023code}).
\item Authors' contributions: ESF and CMH conceptualized this study together. ESF implemented, carried out, and analyzed the simulation results in this paper and was also responsible for original manuscript preparation. CMH analyzed simulation results and provided manuscript editing and review.
\end{itemize}

\noindent

\bibliographystyle{unsrt}
\bibliography{biblio}

\begin{thebibliography}{10}

\bibitem{frizzell2023simulation}
E.S Frizzell and C.M. Hartzell.
\newblock Simulation of lateral impulse induced inertial dilation at the surface of a vacuum-exposed granular assembly.
\newblock {\em Gran. Matt.}, 25(4):75, 2023.
\newblock \url{https://doi.org/10.1007/s10035-023-01363-6}.

\bibitem{nesterenko2013dynamics}
V.F. Nesterenko.
\newblock {\em Dynamics of heterogeneous materials}.
\newblock Springer Science \& Business Media, 2013.
\newblock \url{https://doi.org/10.1007/978-1-4757-3524-6}.

\bibitem{bandfield2014lunarCS}
J.L. Bandfield, E.~Song, P.O. Hayne, B.D. Brand, R.R. Ghent, A.R. Vasavada, and D.A. Paige.
\newblock Lunar cold spots: Granular flow features and extensive insulating materials surrounding young craters.
\newblock {\em Icarus}, 231:221--231, 2014.
\newblock \url{https://doi.org/10.1016/j.icarus.2013.12.017}.

\bibitem{nesterenko1984propagation}
V.F. Nesterenko.
\newblock Propagation of nonlinear compression pulses in granular media.
\newblock {\em J. Appl. Mech. Tech. Phys. (Engl. Transl.); (United States)}, 24(5), 1984.
\newblock \url{https://doi.org/10.1007/BF00905892}.

\bibitem{manciu2001impulse}
M.~Manciu, S.~Sen, and A.J. Hurd.
\newblock Impulse propagation in dissipative and disordered chains with power-law repulsive potentials.
\newblock {\em Physica D: Nonlin. Phenom.}, 157(3):226--240, 2001.
\newblock url{https://doi.org/10.1016/S0167-2789(01)00302-5}.

\bibitem{takato2012long}
Y.~Takato and S.~Sen.
\newblock Long-lived solitary wave in a precompressed granular chain.
\newblock {\em Europhysics Letters}, 100(2):24003, 2012.
\newblock \url{https://doi.org/10.1209/0295-5075/100/24003}.

\bibitem{jiao2023revisiting}
T.~Jiao, W.~Chen, Y.~Takato, S.~Sen, and D.~Huang.
\newblock Revisiting {N}esterenko’s solitary wave in the precompressed granular alignment held between fixed ends.
\newblock {\em Gran. Matt.}, 25(2):17, 2023.
\newblock \url{https://doi.org/10.1007/s10035-023-01309-y}.

\bibitem{rogers1994location}
A.J. Rogers and C.G. Don.
\newblock Location of buried objects by an acoustic impulse technique.
\newblock {\em Acoustics Australia}, 22:5--5, 1994.

\bibitem{sen1998solitonlike}
S.~Sen, M.~Manciu, and J.D. Wright.
\newblock Solitonlike pulses in perturbed and driven {H}ertzian chains and their possible applications in detecting buried impurities.
\newblock {\em Phys. Rev. E}, 57(2):2386, 1998.
\newblock \url{https://doi.org/10.1103/PhysRevE.57.2386}.

\bibitem{sen2005using}
S.~Sen, T.R. Krishna~Mohan, D.P. Visco~Jr, S.~Swaminathan, A.~Sokolow, E.~Avalos, and M.~Nakagawa.
\newblock Using mechanical energy as a probe for the detection and imaging of shallow buried inclusions in dry granular beds.
\newblock {\em Int. J. of Mod. Phys. B}, 19(18):2951--2973, 2005.
\newblock "\url{https://doi.org/10.1142/S0217979205031997"}.

\bibitem{visco2004impulse}
D.P Visco~Jr, S.~Swaminathan, TR-K. Mohan, A.~Sokolow, and S.~Sen.
\newblock Impulse penetration into idealized granular beds: behavior of cumulative surface kinetic energy.
\newblock {\em Phys. Rev. E}, 70(5):051306, 2004.
\newblock \url{https://doi.org/10.1103/PhysRevE.70.051306}.

\bibitem{sen1996sound}
S.~Sen and R.S. Sinkovits.
\newblock Sound propagation in impure granular columns.
\newblock {\em Phys. Rev. E}, 54(6):6857, 1996.
\newblock \url{https://doi.org/10.1103/PhysRevE.54.6857}.

\bibitem{hasan2016universal}
M.A. Hasan and S.~Nemat-Nasser.
\newblock Universal relations for solitary waves in granular crystals under shocks with finite rise and decay times.
\newblock {\em Phys. Rev. E}, 93:042905, Apr 2016.
\newblock \url{https://doi/10.1103/PhysRevE.93.042905}.

\bibitem{hasan2017basic}
M.A. Hasan and S.~Nemat-Nasser.
\newblock Basic properties of solitary waves in granular crystals.
\newblock {\em J. of the Mech. and Phys. of Solids}, 101:1--9, 2017.
\newblock \url{https://doi.org/10.1016/j.jmps.2017.01.004}.

\bibitem{hasan2018shock}
M.A. Hasan and S.~Nemat-Nasser.
\newblock Shock-induced solitary waves in granular crystals.
\newblock {\em Phys. Rev. E}, 97:022205, Feb 2018.
\newblock \url{https:///doi/10.1103/PhysRevE.97.022205}.

\bibitem{pal2013wave}
R.K. Pal, A.P. Awasthi, and P.H. Geubelle.
\newblock Wave propagation in elasto-plastic granular systems.
\newblock {\em Granular Matter}, 15(6):747--758, 2013.
\newblock \url{https://doi.org/10.1007/s10035-013-0449-1}.

\bibitem{pal2014characterization}
R.K. Pal, A.P. Awasthi, and P.H. Geubelle.
\newblock Characterization of wave propagation in elastic and elastoplastic granular chains.
\newblock {\em Phys. Rev. E}, 89(1):012204, 2014.
\newblock \url{https://doi.org/10.1103/PhysRevE.89.012204}.

\bibitem{sen2017impact}
S.~Sen, T.R. Krishna~Mohan, and M.~Tiwari.
\newblock Impact dispersion using 2{D} and 3{D} composite granular packing.
\newblock {\em KONA Powder and Particle J.}, page 2017014, 2017.
\newblock \url{https://doi.org/10.14356/kona.2017014}.

\bibitem{williams2018lunarCS}
J.-P. Williams, J.L. Bandfield, D.A. Paige, T.M. Powell, B.T. Greenhagen, S.~Taylor, P.O. Hayne, E.J. Speyerer, R.R. Ghent, and E.S. Costello.
\newblock Lunar {C}old {S}pots and {C}rater {P}roduction on the {M}oon.
\newblock {\em J. of Geophys. Res.: Planets}, 123(9):2380--2392, 2018.
\newblock \url{https://doi.org/10.1029/2018JE005652}.

\bibitem{hill2017well}
J.R. Hill and P.R. Christensen.
\newblock Well-preserved low thermal inertia ejecta deposits surrounding young secondary impact craters on {M}ars.
\newblock {\em J. of Geophys. Res.: Planets}, 122(6):1276--1299, 2017.
\newblock \url{https://doi.org/10.1002/2016JE005210}.

\bibitem{venkatraman2023statistical}
J.~Venkatraman, T.~Horvath, T.M. Powell, and D.A. Paige.
\newblock Statistical estimates of rock-free lunar regolith thickness from {D}iviner.
\newblock {\em Planetary and Space Science}, 229:105662, 2023.
\newblock \url{https://doi.org/10.1016/j.pss.2023.105662}.

\bibitem{powell2023high}
T.M. Powell, T.~Horvath, V.L. Robles, J.-P. Williams, P.O. Hayne, C.L. Gallinger, B.T. Greenhagen, D.S. McDougall, and D.A. Paige.
\newblock High-resolution nighttime temperature and rock abundance mapping of the {M}oon using the {D}iviner {L}unar {R}adiometer {E}xperiment with a model for topographic removal.
\newblock {\em J. of Geophys. Res.: Planets}, 128(2):e2022JE007532, 2023.
\newblock \url{https://doi.org/10.1029/2022JE007532}.

\bibitem{kloss2012models}
C.~Kloss, C.~Goniva, A.~Hager, S.~Amberger, and S.~Pirker.
\newblock Models, algorithms and validation for opensource {DEM} and {CFD--DEM}.
\newblock {\em Prog. in Comp. Fluid Dyn., an Int. J.}, 12(2-3):140--152, 2012.
\newblock \url{https://doi.org/10.1504/PCFD.2012.047457}.

\bibitem{sanchez2011simulating}
P.~S{\'a}nchez and D.J. Scheeres.
\newblock Simulating asteroid rubble piles with a self-gravitating soft-sphere distinct element method model.
\newblock {\em The Astrophys. J.}, 727(2):120, 2011.
\newblock \url{https://doi.org/10.1088/0004-637X/727/2/120}.

\bibitem{schwartz2012implementation}
S.R. Schwartz, D.C. Richardson, and P.~Michel.
\newblock An implementation of the {S}oft-{S}phere {D}iscrete {E}lement {M}ethod in a high-performance parallel gravity tree-code.
\newblock {\em Granular Matter}, 14(3):363--380, 2012.
\newblock \url{https://doi.org/10.1007/s10035-012-0346-z}.

\bibitem{tancredi2012granular}
G.~Tancredi, A.~Maciel, L.~Heredia, P.~Richeri, and S.~Nesmachnow.
\newblock Granular physics in low-gravity environments using {D}iscrete {E}lement {M}ethod.
\newblock {\em Mon. Notices of the Roy. Astronom. Soc.}, 420(4):3368--3380, 2012.
\newblock \url{https://doi.org/10.1111/j.1365-2966.2011.20259.x}.

\bibitem{sanchez2016disruption}
P.~S{\'a}nchez and D.J. Scheeres.
\newblock Disruption patterns of rotating self-gravitating aggregates: {A} survey on angle of friction and tensile strength.
\newblock {\em Icarus}, 271:453--471, 2016.
\newblock \url{https://doi.org/10.1016/j.icarus.2016.01.016}.

\bibitem{demartini2019using}
J.V. DeMartini, D.C. Richardson, O.S. Barnouin, N.C. Schmerr, J.B. Plescia, P.~Scheirich, and P.~Pravec.
\newblock Using a {D}iscrete {E}lement {M}ethod to investigate seismic response and spin change of 99942 {A}pophis during its 2029 tidal encounter with {E}arth.
\newblock {\em Icarus}, 328:93--103, 2019.
\newblock \url{https://doi.org/10.1016/j.icarus.2019.03.015}.

\bibitem{zhang2021creep}
Y.~Zhang, P.~Michel, D.C. Richardson, O.S. Barnouin, H.F. Agrusa, K.~Tsiganis, C.~Manzoni, and B.H. May.
\newblock Creep stability of the {DART}/{H}era mission target 65803 {D}idymos {II}. {T}he role of cohesion.
\newblock {\em Icarus}, 362:114433, 2021.
\newblock \url{https://doi.org/10.1016/j.icarus.2021.114433}.

\bibitem{sanchez2022transmission}
P.~S{\'a}nchez, D.J. Scheeres, and A.C. Quillen.
\newblock Transmission of a seismic wave generated by impacts on granular asteroids.
\newblock {\em The Planet. Sci. J.}, 3(10):245, 2022.
\newblock \url{https://doi.org/10.3847/PSJ/ac960c}.

\bibitem{frizzell2023data}
E.S. Frizzell.
\newblock Shock induced dilation, restart files and single run output code, 2023.
\newblock \url{https://doi.org/10.5281/zenodo.7608668}.

\bibitem{hartzell2011role}
C.M. Hartzell and D.J. Scheeres.
\newblock The role of cohesive forces in particle launching on the moon and asteroids.
\newblock {\em Planet. and Space Sci.}, 59(14):1758--1768, 2011.

\bibitem{jia1999ultrasound}
X.~Jia, C.~Caroli, and B.~Velicky.
\newblock Ultrasound propagation in externally stressed granular media.
\newblock {\em Phys. Rev. Lett}, 82(9):1863, 1999.
\newblock \url{https://doi.org/10.1103/PhysRevLett.82.1863}.

\bibitem{mckay1991lunar}
D.S. McKay, G.~Heiken, A.~Basu, G.~Blanford, S.~Simon, R.~Reedy, M.F. Bevan, and J.~Papike.
\newblock The lunar regolith.
\newblock In Lunar {S}ourcebook: {A} user's guide to~the {M}oon, editor, {\em Heiken, G.H. and Vaniman, D.T. and Bevan, M.F.}, volume~7, pages 285--356. Press Synd. of the U. of Cambridge, New {Y}ork, 1991.

\bibitem{Rohatgi2022}
A.~Rohatgi.
\newblock Webplotdigitizer: Version 4.6, 2022.
\newblock \url{https://automeris.io/WebPlotDigitizer}.

\bibitem{mase2009continuum}
G.T. Mase, R.E. Smelser, and G.E. Mase.
\newblock {\em Continuum Mechanics for Engineers}.
\newblock CRC press, 2009.
\newblock \url{https://doi.org/10.1201/9780429174391}.

\bibitem{el2008acoustic}
S.AM. El~Shourbagy, S.~Okeda, and H.-G. Matuttis.
\newblock Acoustic of sound propagation in granular materials in one, two, and three dimensions.
\newblock {\em J. of the Phys. Soc. of Japan}, 77(3):034606--034606, 2008.
\newblock \url{https://doi.org/10.1143/jpsj.77.034606}.

\bibitem{gomez2012shocks}
L.R. G{\'o}mez, A.M. Turner, M.~van Hecke, and V.~Vitelli.
\newblock Shocks near jamming.
\newblock {\em Phys. Rev. Letters}, 108(5):058001, 2012.
\newblock \url{https://doi.org/10.1103/PhysRevLett.108.058001}.

\bibitem{van2013shock}
S.~van~den Wildenberg, R.~van Loo, and M.~van Hecke.
\newblock Shock waves in weakly compressed granular media.
\newblock {\em Phys. Rev. Letters}, 111(21):218003, 2013.
\newblock \url{https://doi.org/10.1103/PhysRevLett.111.218003}.

\bibitem{tell2020acoustic}
K.~Tell, C.~Drei{\ss}igacker, A.C. Tchapnda, P.~Yu, and M.~Sperl.
\newblock Acoustic waves in granular packings at low confinement pressure.
\newblock {\em Rev. of Sci. Instr.}, 91(3):033906, 2020.
\newblock \url{https://doi.org/10.1063/1.5122848}.

\bibitem{awasthi2012propagation}
A.P. Awasthi, K.J. Smith, P.H. Geubelle, and J.~Lambros.
\newblock Propagation of solitary waves in {2D} granular media: {A} numerical study.
\newblock {\em Mech. of Materials}, 54:100--112, 2012.
\newblock \url{https://doi.org/10.1016/j.mechmat.2012.07.005}.

\bibitem{stukowski2009visualization}
A.~Stukowski.
\newblock Visualization and analysis of atomistic simulation data with {OVITO}--the {O}pen {V}isualization {T}ool.
\newblock {\em Modelling and Sim. in Mat. Sci. and Eng.}, 18(1):015012, 2009.
\newblock \url{https://doi.org/10.1088/0965-0393/18/1/015012}.

\bibitem{osti_946741}
C~Rycroft.
\newblock Voro++: {A} three-dimensional voronoi cell library in {C}++.
\newblock 1 2009.
\newblock \url{https://doi.org/10.2172/946741}.

\bibitem{zhang2018rotational}
Y.~Zhang, D.C. Richardson, O.S. Barnouin, P.~Michel, S.R. Schwartz, and R.-L. Ballouz.
\newblock Rotational failure of rubble-pile bodies: {I}nfluences of shear and cohesive strengths.
\newblock {\em The Astrophys. J.}, 857(1):15, 2018.
\newblock \url{https://doi.org/10.3847/1538-4357/aab5b2}.

\bibitem{frizzell2023code}
E.S. Frizzell.
\newblock Shock induced dilation, validation code, 2023.
\newblock \url{https://doi.org/10.5281/zenodo.7608511}.

\bibitem{cole2012particle}
D.M Cole, M.A Hopkins, and L~Taylor.
\newblock Particle descriptions and grain contact laws for {L}unar materials: Their implementation in a discrete element model.
\newblock In {\em Earth and {S}pace 2012: {E}ngineering, {S}cience, {C}onstruction, and {O}perations in {C}hallenging {E}nvironments}, pages 208--217. 2012.
\newblock \url{https://doi.org/10.1061/9780784412190.024}.

\bibitem{carrier1991physical}
W.D. Carrier~III, G.R. Olhoeft, and W.~Mendell.
\newblock Physical properties of the lunar surface.
\newblock In Lunar {S}ourcebook: {A} user's guide to~the {M}oon, editor, {\em Heiken, G.H. and Vaniman, D.T. and Bevan, M.F.}, pages 475--594. Press Syndicate of the University of Cambridge, New {Y}ork, 1991.

\bibitem{colwell2009lunar}
J.E. Colwell, S.R. Robertson, M.~Hor{\'a}nyi, X.~Wang, A.~Poppe, and P.~Wheeler.
\newblock Lunar dust levitation.
\newblock {\em J. of Aero. Eng.}, 22(1):2--9, 2009.
\newblock \url{https://doi.org/10.1061/(ASCE)0893-1321(2009)22:1(2)}.

\bibitem{hanus2017volumes}
J.~Hanus, M.~Viikinkoski, F.~Marchis, J.~Durech, M.~Kaasalainen, M.~Delbo, D.~Herald, E.~Frappa, T.~Hayamizu, S.~Kerr, et~al.
\newblock Volumes and bulk densities of forty asteroids from {ADAM} shape modeling.
\newblock {\em A\&A}, 2017.
\newblock \url{https://doi.org/10.1051/0004-6361/201629956}.

\bibitem{yu2014numerical}
Y.~Yu, D.C. Richardson, P.~Michel, S.R. Schwartz, and R.-L. Ballouz.
\newblock Numerical predictions of surface effects during the 2029 close approach of asteroid 99942 {A}pophis.
\newblock {\em Icarus}, 242:82--96, 2014.
\newblock \url{https://doi.org/10.1016/j.icarus.2014.07.027}.

\bibitem{kovach1973velocity}
R.L. Kovach and J.S. Watkins.
\newblock The velocity structure of the lunar crust.
\newblock {\em The Moon}, 7(1):63--75, 1973.
\newblock \url{https://doi.org/10.1007/BF00578808}.

\bibitem{sutton1970elastic}
G.H. Sutton and F.K. Duennebier.
\newblock Elastic properties of the lunar surface from {S}urveyor spacecraft data.
\newblock {\em J. of Geophys. Res.}, 75(35):7439--7444, 1970.
\newblock \url{https://doi.org/10.1029/JB075i035p07439}.

\bibitem{morgan2018pre}
P.~Morgan, M.~Grott, B.~Knapmeyer-Endrun, M.~Golombek, P.~Delage, P.~Lognonn{\'e}, S.~Piqueux, I.~Daubar, N.~Murdoch, C.~Charalambous, et~al.
\newblock A pre-landing assessment of regolith properties at the {I}n{S}ight landing site.
\newblock {\em Space Science Rev.}, 214:1--47, 2018.
\newblock \url{https://doi.org/10.1007/s11214-018-0537-y}.

\bibitem{holmes2016bending}
M.A.J. Holmes, R.~Brown, P.A.L. Wauters, N.P. Lavery, and S.G.R. Brown.
\newblock Bending and twisting friction models in soft-sphere discrete element simulations for static and dynamic problems.
\newblock {\em Appl. Math. Modelling}, 40(5-6):3655--3670, 2016.
\newblock \url{https://doi.org/10.1016/j.apm.2015.10.026}.

\bibitem{chau2002coefficient}
K.T. Chau, R.H.C. Wong, and J.J. Wu.
\newblock Coefficient of restitution and rotational motions of rockfall impacts.
\newblock {\em Int. J. of Rock Mech. and Mining Sci.}, 39(1):69--77, 2002.
\newblock \url{https://doi.org/10.1016/S1365-1609(02)00016-3}.

\bibitem{wang2020behaviors}
J.~Wang, M.~Zhang, L.~Feng, H.~Yang, Y.~Wu, and G.~Yue.
\newblock The behaviors of particle-wall collision for non-spherical particles: {E}xperimental investigation.
\newblock {\em Powder Tech.}, 363:187--194, 2020.
\newblock \url{https://doi.org/10.1016/j.powtec.2019.12.041}.

\bibitem{goddard1990nonlinear}
J.D. Goddard.
\newblock Nonlinear elasticity and pressure-dependent wave speeds in granular media.
\newblock {\em Proc. of the R. Soc. of Lond. S. A: Math. and Phys. Sci.}, 430(1878):105--131, 1990.
\newblock \url{https://doi.org/10.1098/rspa.1990.0083}.

\bibitem{thompson1979blocky}
T.W. Thompson, W.J. Roberts, W.K. Hartmann, R.W. Shorthill, and S.H. Zisk.
\newblock Blocky craters: Implications about the lunar megaregolith.
\newblock {\em The moon and the planets}, 21:319--342, 1979.
\newblock \url{https://doi.org/10.1007/BF00897360}.

\bibitem{richardson2020modeling}
J.E. Richardson and O.~Abramov.
\newblock Modeling the formation of the lunar upper megaregolith layer.
\newblock {\em The Planet. Sci. J.}, 1(1):2, 2020.
\newblock \url{https://doi.org/10.3847/PSJ/ab7235}.

\bibitem{horz1991lunar}
F.~H{\"o}rz, R.~Grieve, G.~Heiken, P.~Spudis, and A.~Binder.
\newblock Lunar surface processes.
\newblock In {A} user's guide to the~{M}oon Lunar~sourcebook, editor, {\em Heiken, G.H. and Vaniman, D.T. and Bevan, M.F.}, volume~7, pages 61--120. Press Synd. of the U. of Cambridge, New {Y}ork, 1991.

\bibitem{somfai2005elastic}
E.~Somfai, J.-N. Roux, J.H. Snoeijer, M.~Van~Hecke, and W.~Van~Saarloos.
\newblock Elastic wave propagation in confined granular systems.
\newblock {\em Phys. Rev. E}, 72(2):021301, 2005.
\newblock \url{https://doi.org/10.1103/PhysRevE.72.021301}.

\end{thebibliography}

\end{document}